\documentclass[twocolumn]{aastex62}

\usepackage{hyperref}
\usepackage{amsmath}
\usepackage{subcaption}
\usepackage{enumitem}
\usepackage[dvipsnames]{xcolor}

\usepackage{soul}

\newcommand{\HII}{H\tiny{ }\footnotesize{II}\normalsize{ }}

\newcommand{\NII}{[N\tiny{ }\footnotesize{II}\normalsize{] }}

\newcommand{\OIII}{[O\tiny{ }\footnotesize{III}\normalsize{] }}

\newcommand{\Hii}{\ion{H}{2}}
\newcommand{\Nii}{[\ion{N}{2}]}

\newcommand{\Oiii}{[\ion{O}{3}]}
\newcommand{\Sii}{[\ion{S}{2}]}
\newcommand{\Siii}{[\ion{S}{3}]}

\newcommand{\Ha}{H$\alpha$}
\newcommand{\Hb}{H$\beta$}

\newcommand{\kms}{\ifmmode\,{\rm km}\,{\rm s}^{-1}\else km$\,$s$^{-1}$\fi}
\newcommand{\um}{$\mathrm{\mu}$m}
\newcommand{\Msun}{\mathrm{M}_{\sun}}
\newcommand{\sfrunit}{\mathrm{M}_{\sun}~\mathrm{yr}^{-1}}

\defcitealias{Dopita16}{D16}

\graphicspath{{./}{}}

\usepackage{lineno}

\usepackage{etoolbox}
\makeatletter
\patchcmd\linenumberpar{\@LN@parpgbrk}{\penalty\@LN@parpgpen\relax}{}{}

\shorttitle{MSA-3D spatially resolved dust attenuation diversity}
\shortauthors{Bari\v{s}i\'{c} et al.}

\begin{document}


\title{MSA-3D: A Diversity of Dust Attenuation Profiles Across the Epoch of Thin Disk Emergence}

\correspondingauthor{Ivana Bari\v{s}i\'{c}}
\email{ibarisic@ucdavis.edu}

\author[0000-0001-6371-6274]{Ivana Bari\v{s}i\'{c}}
\affiliation{Department of Physics and Astronomy, University of California, Davis, 1 Shields Avenue, Davis, CA 95616, USA}

\author[0000-0001-5860-3419]{Tucker Jones}
\affiliation{Department of Physics and Astronomy, University of California, Davis, 1 Shields Avenue, Davis, CA 95616, USA}

\author[0000-0001-9687-4973]{Naveen Reddy}
\affiliation{Department of Physics \& Astronomy, University of California, Riverside, 900 University Avenue, Riverside, CA 92521, USA}

\author[0000-0001-6919-1237]{Matthew Malkan}
\affiliation{Department of Physics and Astronomy, University of California Los Angeles, 430 Portola Plaza, Los Angeles, CA 90095, USA}

\author[0000-0003-4792-9119]{Ryan Sanders}
\affiliation{Department of Physics and Astronomy, University of Kentucky, 505 Rose Street, Lexington, KY 40506, USA}

\author[0000-0002-6586-4446]{Alaina Henry}
\affiliation{Space Telescope Science Institute, 3700 San Martin Drive, Baltimore, MD 21218, USA}




\author[0000-0003-4804-7142]{Ayan Acharyya}
\affiliation{INAF -- Osservatorio Astronomico di Padova, Vicolo Osservatorio 5, 35122 Padova, Italy}

\author[0000-0001-9742-3138]{Kevin Bundy}
\affiliation{Department of Astronomy \& Astrophysics, University of California, Santa Cruz, 1156 High St, CA 95064, USA}

\author[0000-0001-6703-4676]{Juan M. Espejo Salcedo}
\affiliation{Max-Planck-Institut für extraterrestische Physik (MPE), Giessenbachstr., 85748 Garching, Germany}


\author[0000-0002-3254-9044]{Karl Glazebrook}
\affiliation{Centre for Astrophysics and Supercomputing, Swinburne University of Technology, Hawthorn, VIC 3122, Australia}



\author[0000-0003-2804-0648]{Themiya Nanayakkara}
\affiliation{Sydney Institute for Astronomy, School of Physics, University of Sydney, NSW 2006, Australia}

\author[0000-0002-1527-0762]{Danail Obreschkow}
\affiliation{International Centre for Radio Astronomy Research (ICRAR), M468, University of Western Australia, Perth, WA 6009, Australia}

\author[0000-0002-4430-8846]{Namrata Roy}
\affiliation{Raman Research Institute, Sadashivanagar, C. V. Raman Ave., 560080, Bangalore, India}

\author[0000-0002-1499-6377]{Takafumi Tsukui}
\affiliation{Kavli Institute for the Physics and Mathematics of the Universe (WPI), The University of Tokyo, Kashiwa, Chiba 277-8583, Japan}



\author[0000-0003-0980-1499]{Benedetta Vulcani}
\affiliation{INAF -- Osservatorio Astronomico di Padova, Vicolo Osservatorio 5, 35122 Padova, Italy}

\author[0000-0002-9373-3865]{Xin Wang}
\affiliation{School of Astronomy and Space Science, University of Chinese Academy of Sciences (UCAS), Beijing 100049, China}
\affiliation{National Astronomical Observatories, Chinese Academy of Sciences, Beijing 100101, China}
\affiliation{Institute for Frontiers in Astronomy and Astrophysics, Beijing Normal University, Beijing 102206, China}






\begin{abstract}
We present spatially resolved measurements of dust attenuation and star formation in 18 main-sequence star-forming galaxies at z$\sim$1 from the MSA-3D survey, obtained by mapping the Balmer emission lines at $\sim$1~kpc resolution with \textit{JWST}/NIRSpec's MSA in a slit-stepping strategy.  
We investigate the diversity of radial attenuation profiles, and how the spatial variation affects attenuation and star formation rates (SFR) derived from single-aperture measurements. 
We find a notable diversity among radial attenuation profiles: some galaxies exhibit centrally peaked attenuation, but the majority exhibit flat or even positive radial profiles, with large variation at a fixed stellar mass. 
This diversity may reflect different evolutionary pathways shaped by various mechanisms such as disk settling, merging, and internal processes. 
We examine possible biases arising from single-aperture and integrated measurements and find that, while they can under- or over-estimate attenuation and SFRs for individual galaxies, the sample-averaged trends remain roughly unchanged, with the derived SFRs consistent with the star-forming main sequence, and a small scatter.
From our sample, 
we find a median stellar-to-nebular reddening ratio f = E(B-V)$_{\rm star}$/E(B-V)$_{\rm gas}$ of 0.88 with an interquartile range
of 0.51--0.96, suggesting relatively uniform dust distributions even in intermediate-mass galaxies (stellar masses $\sim 10^9$--$10^{10.5}~\Msun$). Our results highlight the importance of spatially resolved attenuation measurements for accurately tracing star formation and understanding the evolving dust geometry in galaxies during a critical epoch of morphological transformation.
\end{abstract}




\section{Introduction} \label{sec:intro}

The \Ha\ recombination line, produced by ionized hydrogen around massive O and B stars, is a widely used tracer of recent star formation over timescales of about $\sim$10 Myr \citep[e.g.,][]{kennicutt1998}. Interpreting the \Ha\ emission requires correction for dust attenuation towards the nebular regions, which can be done using the Balmer decrement (i.e., the observed \Ha/\Hb\ flux ratio). 
However, the distribution of dust within galaxies is non-uniform, ranging from dense birth clouds to the diffuse interstellar medium \citep[e.g.,][]{charlot2000}. 
Especially in spatially resolved observations, the fraction of intrinsic stellar and nebular emission detected depends on which of these regions is being observed.
As a result, patchy or otherwise inhomogeneous dust geometry can cause the attenuation (and by extension the star formation rate, SFR) to be underestimated in spatially-unresolved observations, particularly in central optically thick regions \citep[e.g.][]{calzetti2000, price2014, reddy15, nelson2016_A}.

Most spectroscopic surveys in the local universe rely on single-fiber observations that sample only the central regions of galaxies \citep[e.g. SDSS, GAMA, 6dFGS\footnote{Sloan Digital Sky Survey, Galaxy And Mass Assembly, 6-degree Field Galaxy Survey};][]{york00, driver11, jones09}
, such that attenuation measured from Balmer lines reflects the central region rather than the galaxy as a whole. Local spatially resolved studies \citep[e.g.,][]{greener20} 
show that dust attenuation is typically higher in galaxy centers than in their outskirts. The implication is that applying a central dust correction together with an aperture-based flux correction could bias the inferred total star formation rate \citep[e.g.][]{paramo15, richards16, green17,battisti26}. 
A number of low-redshift studies have tested whether aperture corrections applied to centrally derived measurements reliably recover total SFRs. At $z \lesssim 0.8$, these corrections typically follow the \citet{brinchmann04} prescription, which estimates the total SFR by combining the fiber \Ha\ measurement with a color-based correction for the light outside the fiber, using an empirical relation between optical colors (e.g., g-r or r-i) and specific star formation rate. 
The aperture losses are generally smaller than the intrinsic range of galaxy SFRs, and the corrected measurements recover total SFRs to within $\sim$0.2 dex for unbiased samples \citep{richards16}. 
Aperture corrections are thus broadly reliable in a statistical sense at low redshift, although they recover only an average correction, and cannot capture the resolved attenuation structure within individual galaxies. 
Integral field spectroscopy (IFS), which spatially maps nebular and continuum emission across galaxies, provides a direct way to characterize their resolved properties, and thus test how spatial variations in attenuation affect integrated SFR measurements.

Spatially varying attenuation in galaxies may lead to different biases as a function of redshift, affecting our understanding of dust obscuration and SFR across cosmic time \citep[e.g.,][]{madau2014,hegde2026}. Resolved studies have again shown that galaxies at $z \sim 1$--2 commonly exhibit enhanced central attenuation \citep[e.g.,][]{nelson2016_A, tacchella2018}, yet most spectroscopic studies at $z \gtrsim 1$ rely on single-slit observations. Moreover, there are now large data sets from both ground-based instruments with $\sim$1\arcsec\ seeing and aperture size \citep[e.g.,][]{reddy15} and from JWST/NIRSpec with 0\farcs2 slit width \citep[e.g.,][]{shapley2023}, which may have systematically different biases from aperture effects \citep[e.g.,][]{kewley05, richards16, battisti26}. This motivates a direct spatially resolved assessment of attenuation and aperture effects at intermediate redshifts, which we present in this work. 
We note that much of the current knowledge about spatially varying dust attenuation at $z\gtrsim1$ is based on either composite low-resolution grism spectroscopy \citep{nelson2016_A, nelson2016_B} or photometric measurements \citep[e.g.,][]{tacchella2018}. These studies provide valuable insight into average radial attenuation profiles, but limited information about the sample diversity.
In this paper, we characterize the diversity of attenuation profiles in individual galaxies at z$\sim$1 and assess how spatial variations in dust distribution affect inferred star formation rates and aperture-based measurements. This analysis provides a foundation for future studies of the resolved distribution of star formation, which offer direct insight into the buildup of stellar mass and the onset of quenching -- i.e., the drivers of morphological diversity.

The structure of this paper is as follows. In Section~\ref{sec:Sample} we briefly describe the MSA-3D survey and the sample selection for this study. Section~\ref{sec:methods} introduces the measurement methods and apertures used. In Sections~\ref{sec:results} and \ref{sec:discussion}, we examine and discuss how spectroscopic aperture size affects measured dust attenuation and SFR, using both fixed apertures and radial profiles. The results are summarized in Section~\ref{sec:summary}.
Throughout this work we adopt a flat $\Lambda$CDM cosmology with H$_0 = 69.6$~km\,s$^{-1}$\,Mpc$^{-1}$ and $\Omega_M = 0.286$ \citep{bennett2014}.

\begin{figure}[ht]
    \centering
    \includegraphics[width=\linewidth]{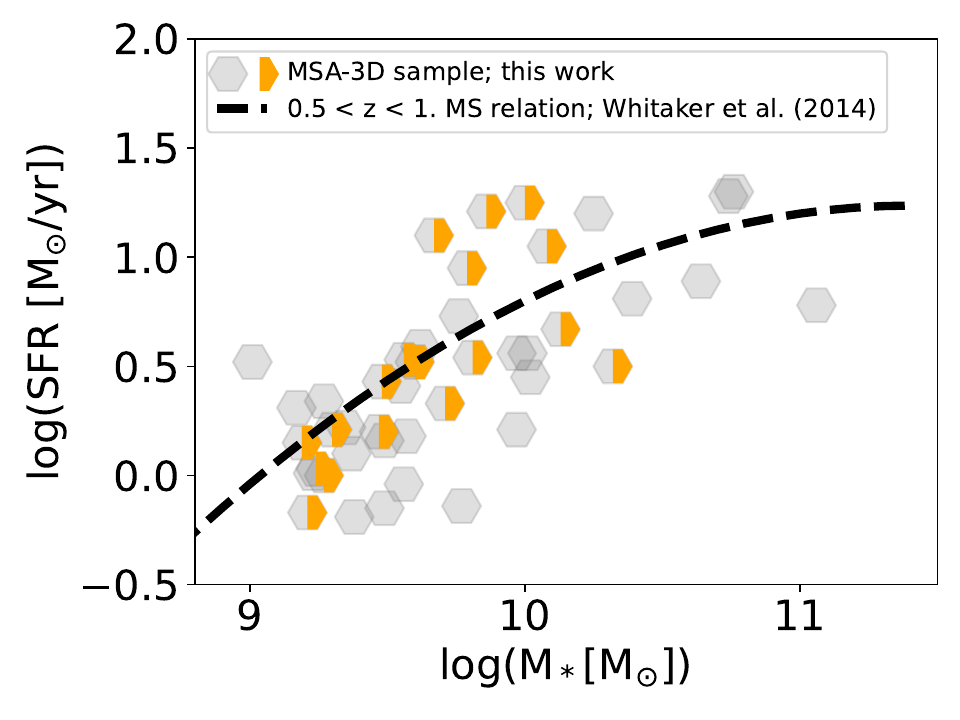}
    \caption{The MSA-3D sample (JWST-GO-2136; grey hexagons) is composed of typical star-forming galaxies within the redshift range $z=0.5$--1.7 with stellar masses $\gtrsim 10^9~\Msun$. Here we show their star formation rates and stellar masses, which generally lie along the star forming main sequence relation from \cite{whitaker2014}, shown as the black dashed line. The subset of galaxies with \Ha\ and \Hb\ coverage which are studied in this work (grey+orange symbols) is broadly representative of the MSA-3D sample.
    }
    \label{fig:sample}
\end{figure}

\begin{figure*}[ht]
    \centering
    \hspace{0.3cm}
    \includegraphics[width=0.95\linewidth]{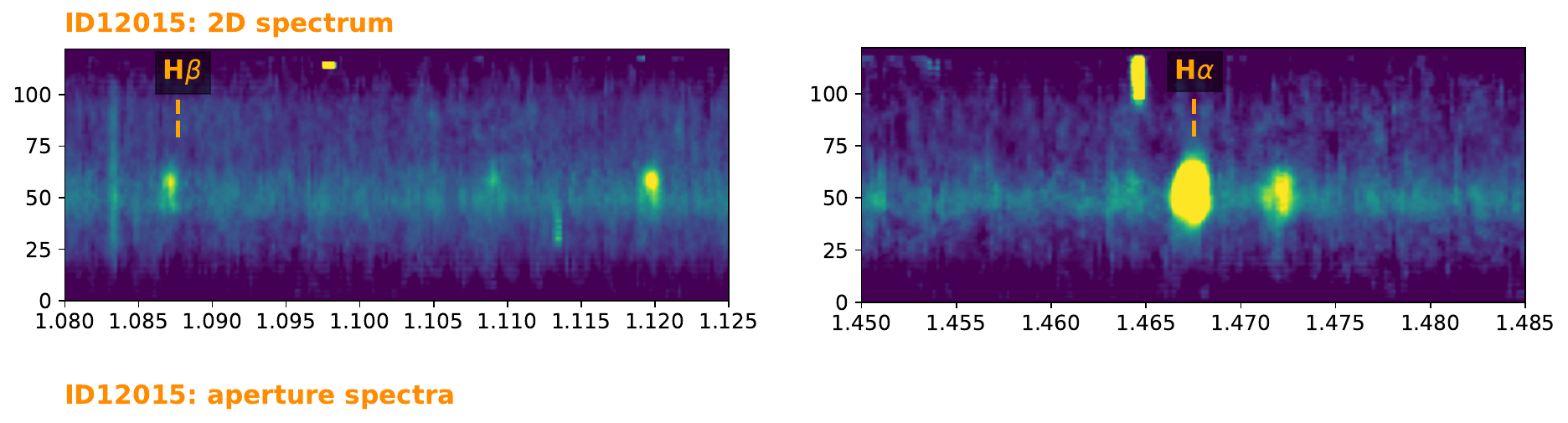}
    \includegraphics[width=\linewidth]{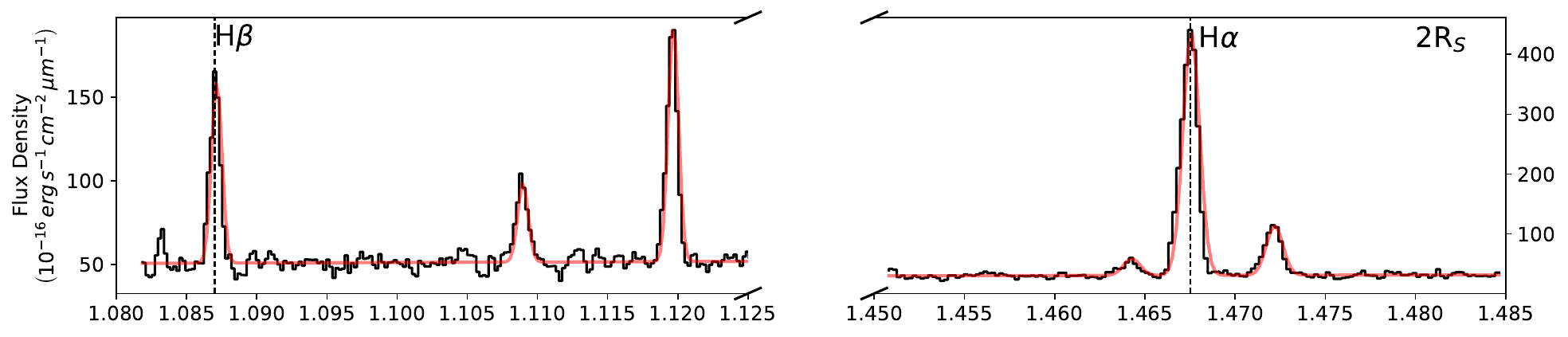}
    \includegraphics[width=\linewidth]{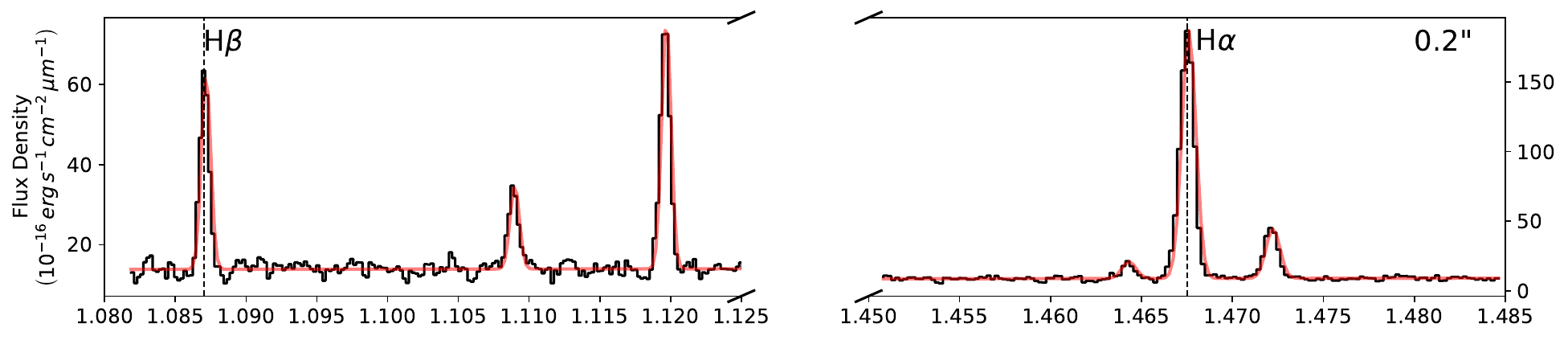}
    \includegraphics[width=\linewidth]{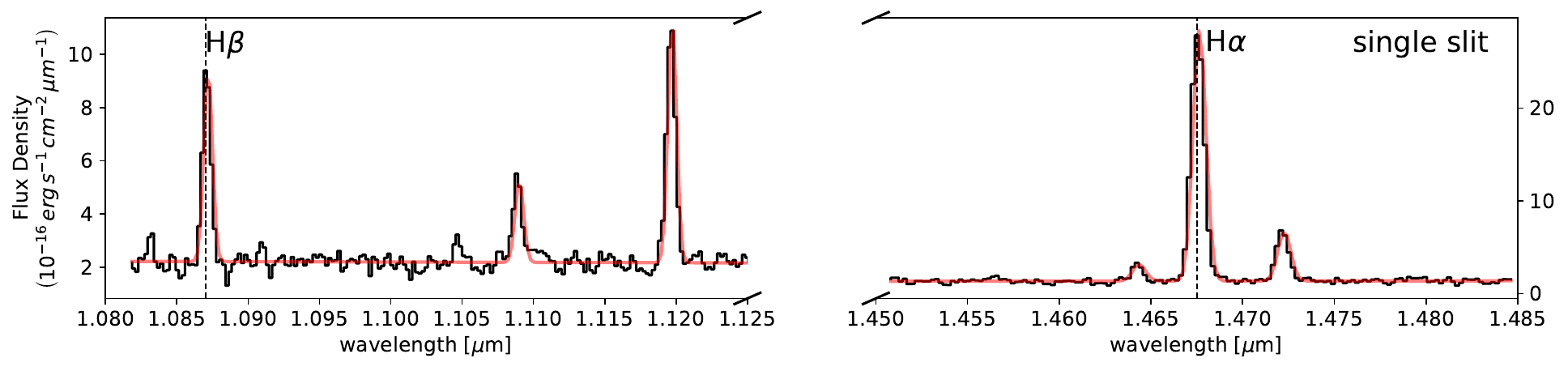}
    \caption{
    Spectra extracted from various apertures for an example galaxy (ID 12015). The top panels show a 2-D spectral slice, revealing spatially extended structure across the galaxy. Lower panels show spectra extracted from the 2$R_S$ aperture (top row), central 0\farcs2 (middle), and single MSA slitlet (bottom). Each row shows the region around \Hb\ + \Oiii, and \Ha\ + \Nii. Spectra are shown in black with best-fit emission line profiles in red. Dashed vertical lines show the observed wavelengths of \Hb\ and \Ha\ at the adopted systemic redshift. The strong Balmer emission provides measurements of nebular attenuation and SFR on sub-galactic scales.}
    \label{fig:emission_lines}
\end{figure*}

\section{Sample and Observations}
\label{sec:Sample}

This paper utilizes data from the JWST Cycle 1 program GO-2136, part of the MSA-3D survey. This program obtained spatially and spectrally resolved pseudo-integral field spectroscopy (IFS) using the NIRSpec Micro-Shutter Assembly (MSA) with a slit-stepping strategy. The survey design details, sample selection, and methodology for constructing IFS cubes using MSA with slit-stepping are described in \cite{barisic2025} \citep[see also][for complementary science results from the MSA-3D Cycle 1 program]{ju2025, ju26, roy26, espejo26}. Here we briefly summarize the key details. 

The program targeted rest-frame optical diagnostic emission lines for a sample of 43 star-forming galaxies within the redshift range $0.5 < z < 1.7$. The sample was selected to have stellar masses $M_{\star} > {10^9}~\Msun$ and star formation rates SFR~$>0.6~\sfrunit$, with observed galaxies generally lying near the star-forming main sequence (Figure~\ref{fig:sample}). 
The sample was observed with a single MSA configuration in the Extended Groth Strip (EGS) field, a region rich in ancillary multi-wavelength photometric and spectroscopic data from surveys such as CANDELS \citep{koekemoer2011}, 3D-HST \citep{momcheva2016}, CEERS \citep{finkelstein2023}, DEEP2/DEEP3 \citep{newman2013}, and MOSDEF \citep{kriek2015}. These ancillary data enabled selection of targets with known redshift, stellar mass, and SFR. 

Each target galaxy was observed with high spectral resolution (R$\sim$2700) using the G140H/F100LP grating and filter combination. The wavelength range is 0.97--1.82~\um, although the wavelength coverage for individual galaxies varies depending on the slit position. Each galaxy was observed with 3--5 slitlets, resulting in a total field of view of 1\farcs8$\times$(2\farcs0--3\farcs0) using a 7$\times$9 pointing slit-stepping dither pattern (see \citealt{barisic2025} for observing pattern details). 
This setup enables the measurement of key rest-frame optical diagnostic lines, including \Hb, \Oiii, \Ha, \Nii, \Sii, and \Siii, depending on the redshift and specific wavelength coverage for each target \citep{barisic2025}. 
For the analysis in this paper we require spectral coverage of both \Ha\ and \Hb, which is available for 18 galaxies spanning the redshift range $1.03 < z < 1.68$. 
From these lines we measure Balmer decrement values in various spatial apertures to determine dust attenuation and dust-corrected star formation rates.

Figure~\ref{fig:sample} presents the full galaxy sample (gray hexagons) and highlights the sub-sample with suitable wavelength coverage used in this study (orange half-hexagons; Table~\ref{tab:sfr_table}). 
Here we adopt total stellar masses and SFRs from the 3D-HST survey catalog \citep{skelton2014,momcheva2016}, as reported in \cite{barisic2025}. 
The selected galaxies 
span the star-forming ``main sequence'' \citep[e.g.,][]{whitaker2014} at $z\sim1$, with stellar masses ranging from $\sim 10^9$--$10^{10.5}~\Msun$.

\section{Methods}
\label{sec:methods}

For this work we are interested in the attenuation and star formation rate within various spatial apertures of our target galaxies. For each aperture of interest, we extract a 1-D spectrum by summing the spectra from all spaxels within the aperture. In this section we describe the measurement methods, followed by the specific apertures used in the subsequent analysis.

\subsection{Emission line measurements}

We extract the 
\Ha\ and \Hb\ line fluxes using Gaussian profile fits. Following \cite{barisic2025} we use a three-component Gaussian model along with a linear continuum to jointly fit neighboring emission lines: \Ha\ together with the \Nii\ doublet, and \Hb\ with the \Oiii\ doublet. 
Each Gaussian component shares a common intrinsic velocity width and redshift, while individual line fluxes are treated as free parameters. 
The \NII and \OIII doublet flux ratios are fixed to their theoretical values of 2.942 and 2.984\footnote{Doublet flux ratios were computed using PyNeb v1.1.15, assuming $T_e = 10,000\ K$ and $n_e = 10^{2}\ cm^{-3}$.}, respectively. 
Figure~\ref{fig:emission_lines} shows examples of Gaussian emission line fits to the spectra extracted from different apertures of a single galaxy in our sample. 
The redshifts and velocity widths of both line complexes are generally in good agreement, confirming that \Ha\ and \Hb\ trace the same physical regions.
The median fitted FWHM (observed frame) is 8.6~\AA\ for the \Ha+\Nii\ complex and 7.3~\AA\ for the \Hb+\Oiii\ complex (median observed FWHM ratio of $1.16\pm0.08$). 
This corresponds to an intrinsic FWHM of 6.8~\AA\ and 5.1~\AA, accounting for instrument resolution of 5.2~\AA\ FWHM. The intrisic median ratio of 1.34 is consistent with the expected value 1.31-1.35 for \Ha\ relative to \Hb+\Oiii, with median intrinsic line widths corresponding to $\sim$150~\kms\ FWHM.

\subsection{Balmer decrement, reddening, and attenuation}
\label{sec:attenuation}

Under the assumption of Case B recombination, an electron temperature of $10^4$~K, and electron density of $10^2$~cm$^{-3}$, the theoretically predicted intrinsic Balmer decrement value is \Ha/\Hb~$=2.86$ \citep{storey1995, osterbrock2006} which we adopt in this work. 
These conditions are typical of \Hii\ regions, and we note that the Balmer decrement has relatively weak sensitivity to temperature 
and density across the ranges found in \Hii\ regions. 
Deviations from this intrinsic value are predominantly due to attenuation and reddening from dust, with higher values corresponding to higher attenuation \citep[e.g.,][]{baker1938,osterbrock2006}.

In this paper we derive the nebular color excess E(B-V) from the observed Balmer decrement using the \cite{cardelli89} reddening curve, and compare it with the color excess derived from stellar continuum photometry using the \cite{calzetti2000} attenuation curve (discussed in Section~\ref{sec:ebv_disc}).
We note that the nebular color excess E(B–V) is relatively insensitive to the assumed dust curve due to the similarity of attenuation curves in the optical regime. 
The nebular reddening E(B-V) is related to the measured Balmer decrement as
\begin{equation}
\begin{split}
    E(B-V) & =  \frac{E(H\beta - H\alpha)}{(k_{H\beta} - k_{H\alpha})} \\
    & = \frac{2.5}{{(k_{H\beta} - k_{H\alpha})}} \log_{10} \frac{(H\alpha/H\beta)_{obs}}{2.86}
\end{split}    
\end{equation}
where $k_\lambda$ is the reddening curve value at the corresponding wavelength $\lambda$, and 2.86 is the intrinsic flux ratio of \Ha/\Hb. 

The nebular reddening E(B–V), or similarly E(\Hb-\Ha), provides a measure of the dust column along the line-of-sight (under the idealized assumption of a uniform foreground screen) toward the emission region probed by the observation. 
We can convert the color excess to attenuation using the relation $\mathrm{A_\lambda = k_\lambda \times E(B-V)}$ 
where $k_{\lambda}$ is determined by the adopted reddening curve. 
The conversion factors for the \citet[][A(\Ha) and A(\Hb)]{cardelli89} and \citet[][stellar A(V)]{calzetti2000} reddening curves are: 
\begin{align}
    \nonumber A(H\alpha) = 2.54 \times E(B-V)  \\ 
    \nonumber A(H\beta) = 3.61 \times E(B-V)  \\  
    \nonumber A(V) = 4.05 \times E(B-V)   
\end{align}
Attenuation of \Ha\ and \Hb\ are required for deriving intrinsic star formation rates from the nebular emission lines, while the V-band attenuation A(V) can be directly compared with that inferred from stellar population synthesis modeling.

\subsection{Star formation rates from \Ha}
\label{sec:sfr_halpha}

The dust-corrected \Ha\ flux provides a measurement of the star formation rate, as \Ha\ traces ionizing radiation from young, massive stars. 
We correct \Ha\ flux for attenuation (Section~\ref{sec:attenuation}), convert this flux to intrinsic luminosity, and compute the SFR via
\begin{equation}
    \mathrm{SFR = 4.68 \times 10^{-42} \, \frac{L(H_{\alpha})}{erg\,s^{-1}} \, M_{\odot}\,yr^{-1}}. 
\end{equation}
This is the calibration from \citet{kennicutt1998}, updated with a \cite{chabrier2003} initial mass function.

\subsection{Apertures and radial profiles}
\label{sec:apertures}

\begin{figure}[ht]
    \centering
    \includegraphics[width=\linewidth]{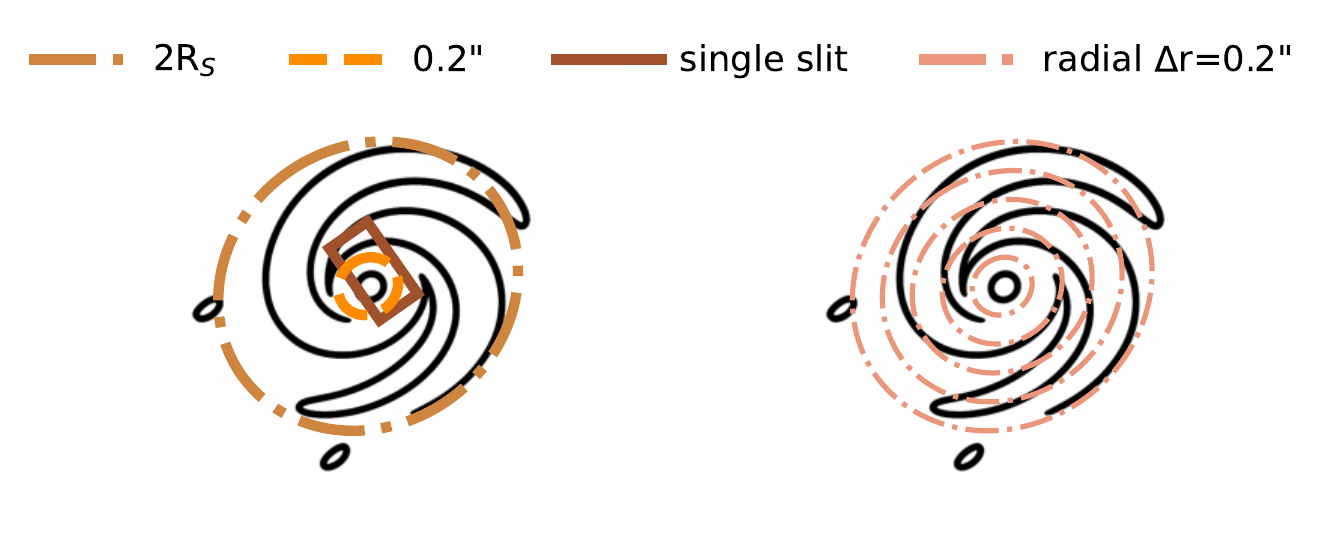}
    \caption{Illustration of two approaches we use to measure Balmer decrement values from MSA-3D IFS data.
    {\bf{Left:}} Three single apertures corresponding to galaxy-integrated (2$R_S$; light brown dash-dotted ellipse), central (0\farcs2; orange ellipse), and single-slit (NIRSpec MSA-like; dark brown rectangle) measurements. Similar apertures are often used in various studies. 
    {\bf{Right:}} Concentric radial apertures of 0\farcs2 width ($\sim$1.7~kpc), used to quantify the radial Balmer decrement and SFR profiles across the galaxy.
    }
    \label{fig:cartoon}
\end{figure}

We aim to assess uncertainty and possible biases in galaxies' SFR and attenuation from single-aperture measurements which are commonly used in the literature. To do so, we analyze spectra extracted from various apertures, as well as trends with galactocentric radius. 
Many apertures considered in this work are defined using measurements of the galaxy center, axis ratio (b/a, which is related to inclination in the case of disk-dominated galaxies), and position angle, reported in Table~\ref{tab:sfr_table}. 
These structural parameters are obtained using GALFIT, assuming a single Sersic profile for each target. The fitting is performed on JWST/NIRCam F444W images from the CEERS survey \citep{finkelstein2023}, and on HST/WFC3 F160W images for targets with no NIRCam coverage. 
The specific apertures we use are illustrated schematically in Figure~\ref{fig:cartoon} and defined as follows.

\begin{enumerate}
    \item Single aperture: we consider three cases for each galaxy, motivated by the common use of similar apertures in the literature. 
    \begin{enumerate}[label=(\alph*)]
        \item 2 scale radii ($2 \mathrm{R_S}$) aperture\\
    	This encompasses 2 scale radii, where the scale radius R$_S$ is defined as a radius at which the galaxy’s surface brightness drops to $1/e$ of its central value. 
        We use an elliptical aperture with the axis ratio and position angle measured from direct imaging (Table~\ref{tab:sfr_table}). Assuming an exponentially declining surface brightness profile, which is typical for star-forming disk-dominated galaxies,  
        this aperture captures the majority of the galaxy’s total light and serves as a proxy for integrated, galaxy-wide flux. 
        We adopt a correction factor of $1.7\times$ from the flux within $2 \mathrm{R_S}$ to the total flux, appropriate for an exponential surface brightness profile. This factor is relevant only for the comparison with photometrically-derived SFR values, and does not affect other aspects of the analysis.
        \item 0\farcs 2 aperture\\
            This is an elliptical aperture with semi-major axis 0\farcs2. 
            The center, position angle, and axis ratio are the same as used above. This aperture is used to probe the dust attenuation in the central region. For the redshifts of our targets, 0\farcs2 corresponds to approximately 1.7~kpc. 
            We verified that the 0\farcs2 aperture does not underestimate the inferred central attenuation: using a smaller 0\farcs1 semi-major axis aperture yields consistent central Balmer decrement values, with no systematic increase in attenuation towards the center.
        \item single MSA slitlet aperture \\
            We extract the spectrum corresponding to a single NIRSpec MSA slitlet, specifically the central pointing of our slit-stepping dither pattern, which has a fixed width of 0\farcs2 and length of 0\farcs46.  
            This aperture provides an assessment of properties derived from standard multi-object spectroscopic surveys with JWST/NIRSpec's MSA. 
            As with single-slit MSA observations, the central slitlet used for extraction does not perfectly align with the galaxy center. This aperture thus provides a reliable basis for assessing the performance of standard MSA observations. 
    \end{enumerate}
    \item Radial attenuation profiles: 
    we construct elliptical annular bins of width 0\farcs2 along the major axis (reaching $\gtrsim$2R$_S$), corresponding to $\sim$1.7~kpc in deprojected radius. 
    Deprojection assumes a thin, circular disk ($\sin(i)=b/a$), which may be less accurate for galaxies with substantial bulges or unsettled disks. 
    We use the same center, position angle, and axis ratio measured from direct imaging to derive spatially resolved attenuation and star formation rate profiles. 
\end{enumerate}

For each aperture or radial bin, we extract total \Ha\ and \Hb\ fluxes from Gaussian fits to the resulting 1D spectra (e.g., Figure~\ref{fig:emission_lines}). We apply the framework described previously in this section to measure the nebular reddening E(B$-$V), attenuation A(H${\alpha}$) and A(H${\beta}$), and attenuation-corrected star formation rates. SFR and A(H${\alpha}$) values are reported for each of the adopted single apertures in Tables~\ref{tab:sfr_table} and \ref{tab:att_table}, respectively.

\subsection{Balmer absorption correction}

Accurate \Ha\ and \Hb\ fluxes are essential for measuring dust attenuation and SFRs. Both lines are affected by underlying stellar Balmer absorption, which reduces the measured emission line fluxes, particularly for \Hb. If unaccounted for, stellar Balmer absorption potentially leads to an overestimation of the Balmer decrement and consequently of the inferred dust attenuation, and an underestimate of the corresponding SFRs. Here we assess the stellar Balmer absorption using the observed spectra. Individual galaxies typically have insufficient signal/noise in the stellar continuum to determine the absorption profiles. We therefore stack the integrated (within $2 \mathrm{R_S}$) spectra of the entire sample, and fit the resulting high-S/N stacked spectrum using pPXF \citep{cappellari17} with FSPS models \citep{conroy09, conroy10}. 
Using the stellar continuum model instead of a simple linear continuum results in an increased best-fit emission line equivalent width of 0.61~\AA\ for \Ha\ and 0.57~\AA\ for \Hb, comparable to results from previous studies of $z \sim 1$ star-forming galaxies observed with similar spectral resolution \citep[e.g.,][]{zahid2011}. This establishes the typical absorption correction factors for our sample. 
For comparison, the \Hb\ emission lines in our sample have a median equivalent width of $\sim15$~\AA\ (16th--84th percentile: $7-34$~\AA).
We note that this correction is much smaller than the total stellar absorption, since the emission lines observed with $R \sim 2700$ spectroscopy are narrower than the stellar Balmer absorption. 
Thus, the correction corresponds only to the stellar absorption underlying the narrow emission line.

To account for variation in the stellar absorption among individual galaxies,
and in different parts of each galaxy, we use the best-fit SED templates derived from BAGPIPES \citep{carnall18}.  
While the continuum shape is reproduced well, we find that the SED best-fit template spectra from BAGPIPES systematically overestimate the Balmer absorption and are inconsistent with the stacked spectrum. We define a rescaling factor as the ratio of absorption equivalent widths in the best-fit pPXF model compared to the stacked BAGPIPES SED template spectra. We then scale the SED-based Balmer absorption for each galaxy by this factor to ensure the average stellar absorption strength is matched to the observed stacked spectrum. Ultimately, the stellar Balmer absorption correction increases the emission line fluxes by a median of 1.3\% for \Ha\ and 5.5\% for \Hb. The sample 16--84th percentile scatter is 0.8--3.4\% for \Ha\ and 2.8--12.6\% for \Hb. The derived Balmer decrement decreases modestly by a median of 4\% (16--84th percentile: 2.0-7.4\%). SFR values in Table~\ref{tab:sfr_table} and attenuation A(\Ha) values in Table~\ref{tab:att_table} include this correction. 
In general the stellar absorption corrections are small, and we find no significant difference in the main results of this paper using corrections based on individual galaxy SED templates or a constant mean correction. A constant correction produces a comparable median decrease of $\sim$3\%, and the scatter in the Balmer decrement is not reduced by individual corrections relative to the constant one ($\sigma$: 0.49 vs. 0.48). The corresponding decrease in the derived attenuation is small (0.09~mag in A(\Ha) for the constant, and 0.15~mag for individual corrections), and does not significantly alter our main results.

\section{Results}
\label{sec:results}

In this section, we analyze how the choice of aperture influences measurements of dust attenuation and the resulting star formation rates. We begin by analyzing attenuation derived from three fixed apertures to assess how global trends vary with aperture size. We then examine spatially resolved attenuation using radial profiles, which reveal dust gradients within individual galaxies and highlight the limitations of aperture-based measurements.

\subsection{Single-aperture measurements}
\label{sec:single-aperture}

\begin{figure*}[ht]
    \centering
    \includegraphics[width=\linewidth]{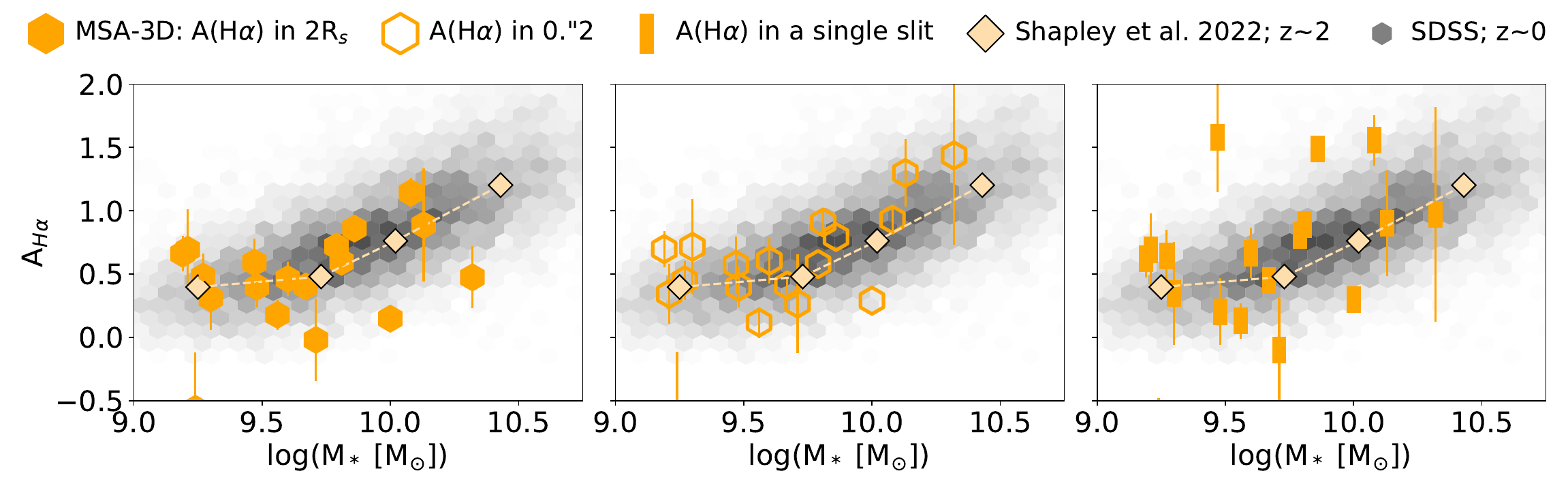}
    \includegraphics[width=\linewidth]{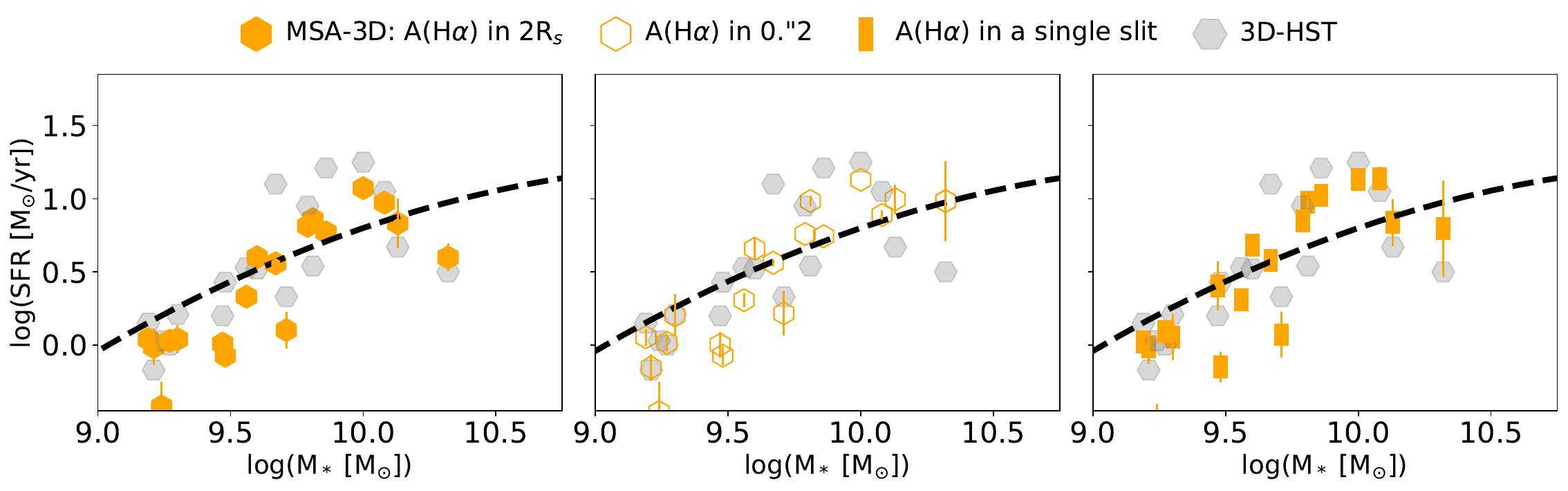}
    \caption{{\bf{Top}}: Attenuation A(\Ha) as a function of stellar mass across the redshift range $z=0$–-2.
    We show the attenuation derived from Balmer decrement measurements using three single apertures: galaxy-wide (2$R_S$; left), central (0\farcs2; middle), and single-slit (MSA-like; right). In all cases, the measured attenuation increases with stellar mass, consistent with trends observed at both $z\sim0$ and $z\sim2$, regardless of the aperture choice. 
    {\bf{Bottom}}: We show SFRs derived from attenuation-corrected H$\alpha$ fluxes measured within a 2$R_S$ aperture, using three aperture-based attenuation corrections: galaxy-wide (2$R_S$; left), central (0\farcs2; middle), and single-slit (MSA-like; right). 
    For comparison, SED-based SFRs from the 3D-HST catalog are shown as grey hexagons. SFRs based on dust-corrected H$\alpha$ fluxes show comparable scatter around the star-forming main sequence (0.28, 0.26, 0.29~dex respectively) relative to SED-based SFRs (0.29~dex). 
    These results illustrate that population-level trends are  recovered even when using small apertures to derive global attenuation corrections.
    }
    \label{fig:attenuation-mass}
\end{figure*}

In Figure~\ref{fig:attenuation-mass} we show the nebular attenuation measured from Balmer decrements in three different apertures described in Section~\ref{sec:apertures}: 2R$_S$ (top left panel), 0\farcs2 (top middle), and the single MSA slitlet (top right).
Previous studies have reported a positive correlation between the Balmer decrement (and thus attenuation) and stellar mass, with more massive galaxies exhibiting higher levels of dust obscuration \citep[e.g.,][]{garn10, dominguez2013, kashino13, reddy15}. It has also been suggested that this relation persists with no significant evolution from the local universe out to at least $z\approx3$--6 \citep{shapley2022,shapley2023,song2026}.

Attenuation measurements for our $z\sim1$ sample confirm this observed redshift-independent positive trend with stellar mass across all apertures considered. These measurements are compared with those from the SDSS \citep[$z\sim0$;][]{abdurrouf2022} and MOSDEF \citep[$z \sim 2$;][]{shapley2022} surveys. SDSS fiber and MOSDEF slit apertures correspond most closely to our 0\farcs2 and 2R$_S$ apertures, respectively. We find however that the observed average trends are robust to aperture differences. The observed scatter in our attenuation–stellar mass relation (standard deviation: 0.3-0.4~mag) is consistent with both the SDSS and MOSDEF surveys, further supporting the observed minimal evolution in this relation from $z\sim0$ to $z\sim2$. 

We note that two galaxies in our sample (IDs 29470, 10863; see Table~\ref{tab:att_table}) have negative nebular attenuation, with A(\Ha)(2R$_S$)$=-0.02\pm0.33$~mag and $-0.56\pm0.44$~mag, respectively. Both values are consistent with zero within the uncertainties, with galaxy 10863 lying $\sim1.3\sigma$ below zero. These negative values are likely due to observational uncertainty, primarily the low S/N of \Hb\, which can result in the measured Balmer decrement below the Case B value of 2.86. While some recent studies interpret Balmer decrements below 2.86 as possible deviations from Case B \citep[e.g.][]{sun25}, others find them consistent when taking measurement uncertainties into account \citep[e.g.][]{matharu23, sandles24, watson26}. With only one galaxy $\sim1.3\sigma$ below zero, we are not able to test the validity of Case B recombination, and we attribute these negative values to measurement uncertainty.

The bottom panels of Figure~\ref{fig:attenuation-mass} present the dust-corrected SFRs, all derived using \Ha\ fluxes measured within the 2R$_S$ aperture. Each panel corresponds to a different attenuation correction, based on Balmer decrement measurements from one of the three apertures (2R$_S$, 0\farcs2, and a single MSA slit). These \Ha-based SFRs are also compared with SED-based SFRs from 3D-HST \citep{momcheva2016}. In all cases, \Ha\ based SFRs exhibit a comparable  
scatter around the star-forming main sequence (0.26-0.29~dex) relative to the SED-based estimates (0.29~dex). 
The corresponding values of dust-corrected SFRs using attenuation derived from each aperture are listed in Table~\ref{tab:sfr_table}.
The consistency of the dust-corrected SFRs across all three attenuation apertures indicates that the sample-wide results are insensitive to aperture choice. This further suggests that the observed non-evolving attenuation--stellar mass relations reported by local and high-redshift surveys are insensitive to differences in the physical regions sampled.

\begin{table*}[htbp]
\centering
\scriptsize 
\begin{tabular}{rrrrrrrrrrr}
\hline\hline
\textbf{ID} & \textbf{RA} & \textbf{DEC} & \textbf{z} & \textbf{PA} & \textbf{b/a} & \textbf{log(M$_\star$)} & \textbf{SFR(SED)} & \textbf{SFR(2R\(_S\))} & \textbf{SFR(0\farcs2)} & \textbf{SFR(slit)} \\
    &     &     &     &    &     &   [M$_{\odot}$]  &   [M$_\odot$/yr] & [M$_\odot$/yr] & [M$_\odot$/yr] & [M$_\odot$/yr] \\
\hline
2145 & 215.069468 & 52.910852 & 1.173166 & 8.01 & 0.35 & 9.19 & 1.41 & 1.09 $\pm$ 0.14 & 1.13 $\pm$ 0.15 & 1.05 $\pm$ 0.14 \\
2465 & 215.070438 & 52.913724 & 1.245019 & 4.69 & 0.28 & 9.30 & 1.62 & 1.10 $\pm$ 0.25 & 1.60 $\pm$ 0.54 & 1.14 $\pm$ 0.42 \\
3399 & 215.042511 & 52.899603 & 1.335943 & 68.29 & 0.30 & 9.81 & 3.47 & 7.24 $\pm$ 0.58 & 9.65 $\pm$ 0.80 & 9.49 $\pm$ 0.84 \\
4391 & 215.067618 & 52.923244 & 1.078873 & -67.99 & 0.92 & 9.48 & 2.69 & 0.85 $\pm$ 0.12 & 0.85 $\pm$ 0.12 & 0.71 $\pm$ 0.17 \\
6199 & 215.045016 & 52.919465 & 1.591923 & -27.39 & 0.49 & 10.00 & 17.78 & 11.82 $\pm$ $\infty$ & 13.47 $\pm$ $\infty$ & 13.55 $\pm$ $\infty$ \\
6430 & 215.013144 & 52.898038 & 1.171503 & 30.00 & 0.34 & 9.79 & 8.91 & 6.52 $\pm$ $\infty$ & 5.74 $\pm$ $\infty$ & 7.06 $\pm$ $\infty$ \\
7561 & 215.060909 & 52.938383 & 1.029531 & -31.89 & 0.32 & 9.21 & 0.68 & 0.97 $\pm$ 0.28 & 0.70 $\pm$ 0.15 & 0.97 $\pm$ 0.25 \\
8365 & 215.059990 & 52.942237 & 1.684560 & -42.65 & 0.50 & 9.56 & 3.39 & 2.14 $\pm$ 0.23 & 2.03 $\pm$ 0.22 & 2.05 $\pm$ 0.26 \\
8512 & 215.049781 & 52.938080 & 1.104036 & 89.64 & 0.69 & 10.32 & 3.16 & 3.97 $\pm$ 0.87 & 9.63 $\pm$ 6.08 & 6.27 $\pm$ 4.74 \\
8576 & 215.059568 & 52.943433 & 1.566937 & 0.16 & 0.82 & 9.60 & 3.31 & 3.99 $\pm$ 0.49 & 4.55 $\pm$ 0.75 & 4.82 $\pm$ 0.87 \\
8942 & 215.009403 & 52.910066 & 1.175502 & 5.40 & 0.74 & 9.86 & 16.22 & 5.89 $\pm$ $\infty$ & 5.54 $\pm$ $\infty$ & 10.49 $\pm$ $\infty$ \\
10502 & 214.985771 & 52.903305 & 1.230851 & 47.13 & 0.59 & 10.13 & 4.68 & 6.79 $\pm$ 2.67 & 9.91 $\pm$ 2.32 & 6.88 $\pm$ 2.53 \\
10863 & 215.055126 & 52.952991 & 1.029679 & 28.92 & 0.62 & 9.24 & 1.07 & 0.38 $\pm$ 0.15 & 0.35 $\pm$ 0.17 & 0.21 $\pm$ 0.13 \\
11225 & 215.041579 & 52.945465 & 1.052296 & -25.91 & 0.40 & 9.67 & 12.59 & 3.62 $\pm$ 0.20 & 3.65 $\pm$ 0.21 & 3.77 $\pm$ 0.22 \\
11944 & 215.036990 & 52.945394 & 1.038734 & 71.53 & 0.46 & 9.27 & 1.00 & 1.07 $\pm$ 0.17 & 1.04 $\pm$ 0.15 & 1.24 $\pm$ 0.24 \\
12015 & 215.032315 & 52.943180 & 1.235711 & -41.57 & 0.67 & 10.08 & 11.22 & 9.40 $\pm$ 0.99 & 7.74 $\pm$ 0.62 & 13.76 $\pm$ 2.52 \\
12071 & 215.021966 & 52.936057 & 1.276201 & -14.81 & 0.30 & 9.47 & 1.58 & 1.02 $\pm$ 0.17 & 1.01 $\pm$ 0.20 & 2.54 $\pm$ 0.99 \\
29470 & 214.968951 & 52.945373 & 1.041457 & -47.02 & 0.64 & 9.71 & 2.14 & 1.27 $\pm$ 0.36 & 1.65 $\pm$ 0.57 & 1.18 $\pm$ 0.43 \\
\hline
\end{tabular}\\
\caption{Sample properties and star formation rate (SFR) measurements. For each target galaxy we list the source ID, RA and Dec coordinates (J2000), redshift ($z$), position angle (PA), and b/a axis ratio which are used to define various apertures.}
\label{tab:sfr_table}
\end{table*}

\subsection{Radial attenuation profiles}

We derive radial attenuation profiles for each galaxy by measuring the integrated \Ha/\Hb\ ratio in radial bins of 0\farcs2 (Figure~\ref{fig:cartoon}). These profiles reveal a diversity in dust attenuation gradients across the sample, ranging from negative (i.e., centrally peaked attenuation) to flat or positive gradients (i.e., centrally reduced attenuation). Four representative examples illustrating this variety are shown in the left panels of Figure~\ref{fig:radial_trends} (with the full sample shown in Appendix~\ref{sec:attenuation_gradients_fullsample}). 
For each of these galaxies, we also show the corresponding surface brightness profiles of \Ha\ and \Hb\, before and after correcting for dust attenuation (middle panels, Figure~\ref{fig:radial_trends}).

\begin{figure*}[ht]
    \centering 
    \includegraphics[width=\linewidth]{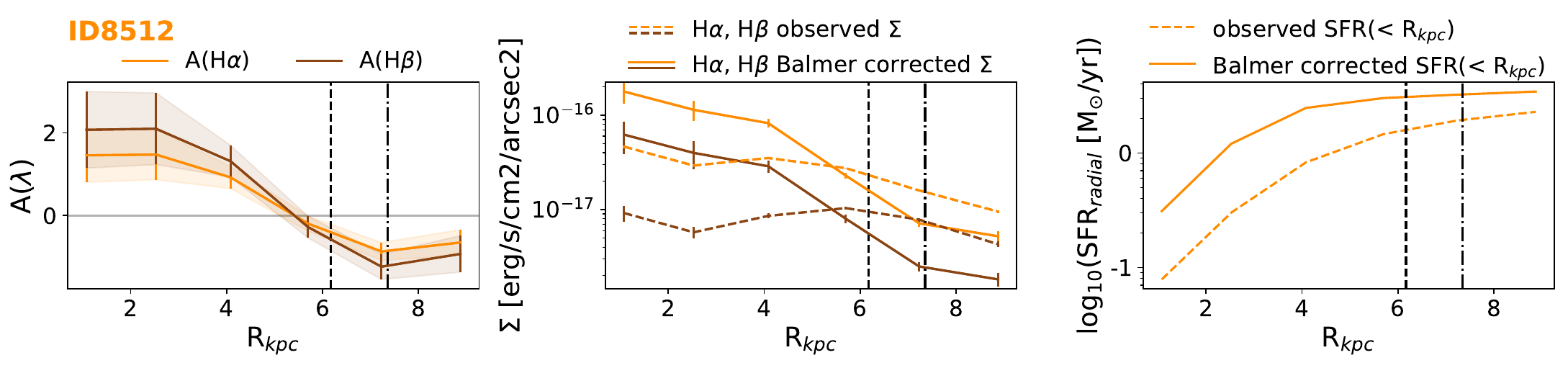}
    \includegraphics[width=\linewidth]{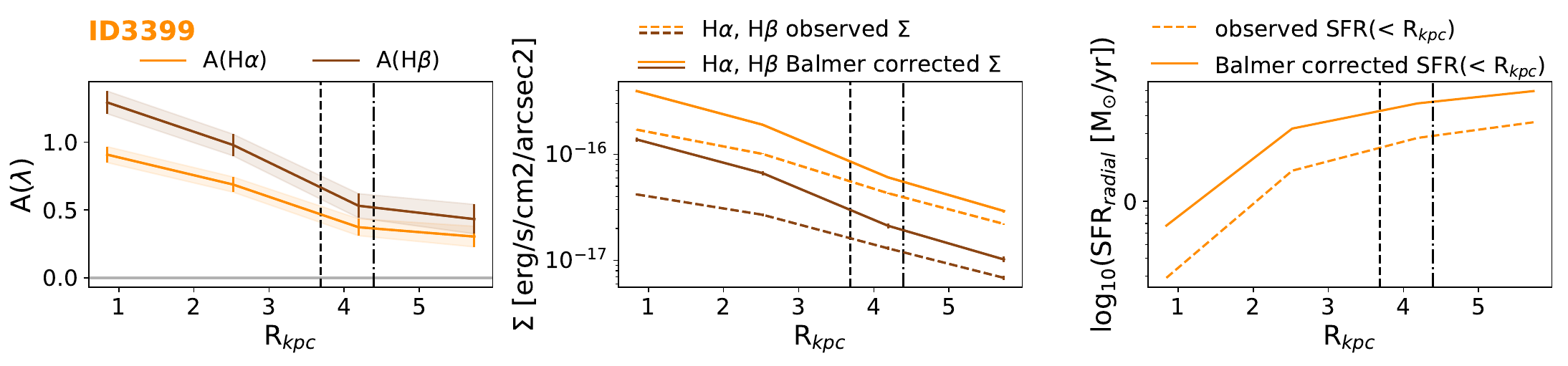}
    \includegraphics[width=\linewidth]{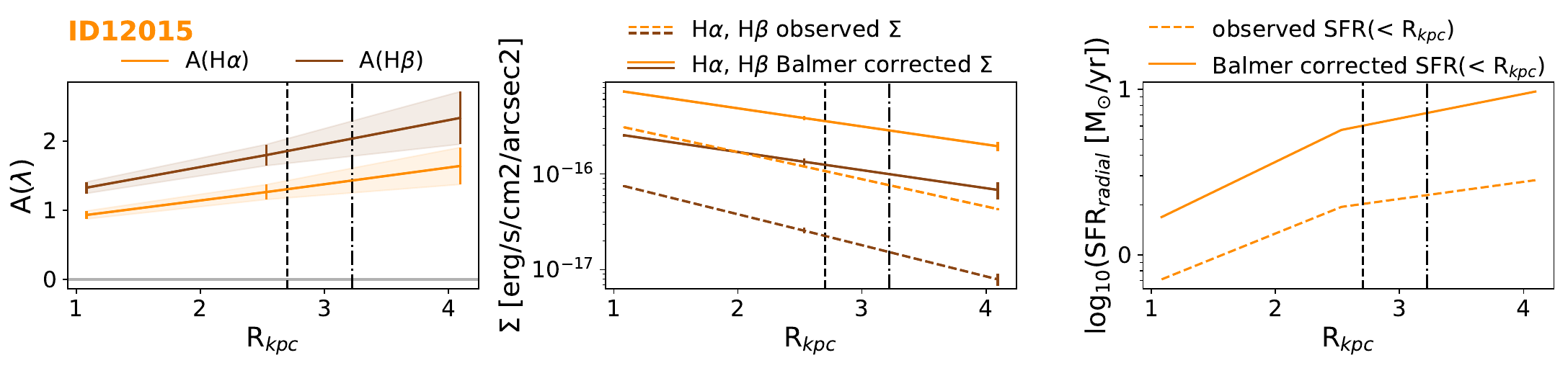}
    \includegraphics[width=\linewidth]{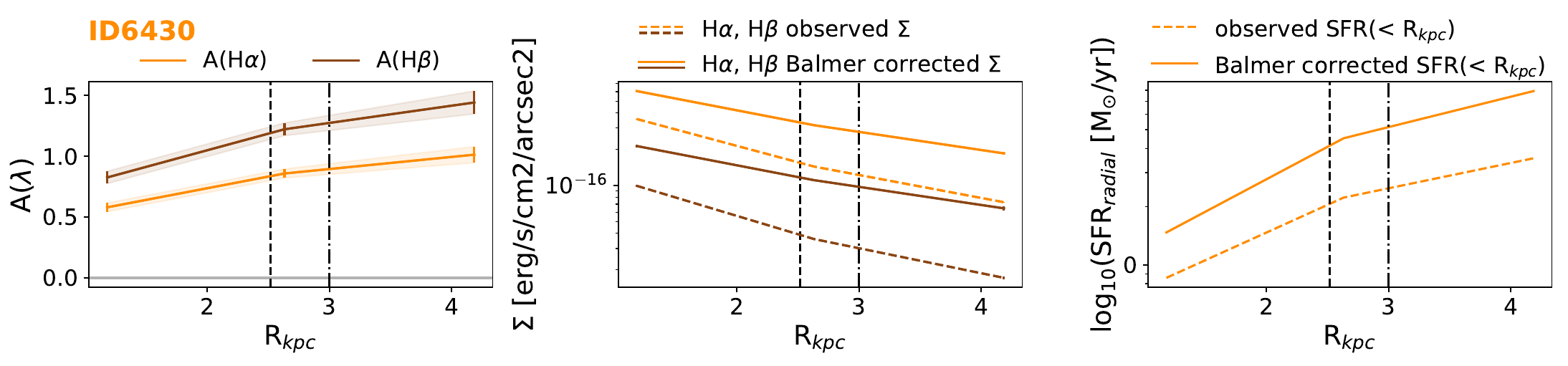}
    \caption{Radial profiles of four example galaxies in our sample. {\bf{Left:}} Radial attenuation profiles for \Ha\ and \Hb\, derived from the Balmer decrement measured in each radial bin, with $1\sigma$ uncertainty propagated from the Balmer decrement uncertainty in each bin. Example profiles of four representative galaxies are shown,
    highlighting the diversity of attenuation gradients observed across the sample, ranging from negative (top) to relatively flat and positive (bottom) trends.  
    {\bf{Middle:}} Surface brightness profiles of the observed (dashed) and dust-corrected (solid) \Ha\ and \Hb\ emission. The profiles are consistent with an exponential profile characteristic of disk galaxies, regardless of the attenuation gradient.  
    {\bf{Right:}} Cumulative radial star formation rate profiles enclosed in a given radius based on \Ha\ flux, shown both before (dashed) and after (solid) correction for dust attenuation. Dashed lines show the effective radius (R$_{eff}$), while dash-dotted lines represent two scale radii (2R$_S$).
    }
    \label{fig:radial_trends}
\end{figure*}

Regardless of the attenuation gradient, the dust-corrected profiles are broadly consistent with exponentially declining surface brightness profiles typical of disk galaxies. 
We note that a stronger central attenuation produces a steeper corrected surface brightness profile (and smaller scale radius), while a lower central attenuation yields a shallower profile.

We also show the cumulative radial profiles of SFR, calculated from the \Ha\ flux enclosed within each 0\farcs2 annular bin. These profiles are shown both before and after correction for dust attenuation.  
These cumulative profiles quantify the fraction of the total \Ha\ flux, and hence SFR, enclosed as a function of radius. They show that the 2R$_S$ aperture captures the majority (see Section~\ref{sec:totalSFR}) 
of the total dust-corrected star formation, supporting its use as a proxy for the integrated SFR.

While the sample-averaged trends recovered from single-aperture corrections remain robust (Section~\ref{sec:single-aperture}, Figure~\ref{fig:attenuation-mass}), Figure~\ref{fig:radial_trends} shows that radial attenuation profiles vary substantially among individual galaxies. Integrated aperture measurements are thus sufficient for statistical studies of galaxy samples, while spatially resolved measurements are required to fully characterize the structure of interstellar medium and SFR within individual sources.

\begin{figure*}[ht]
    \centering
    \includegraphics[width=0.45\linewidth]{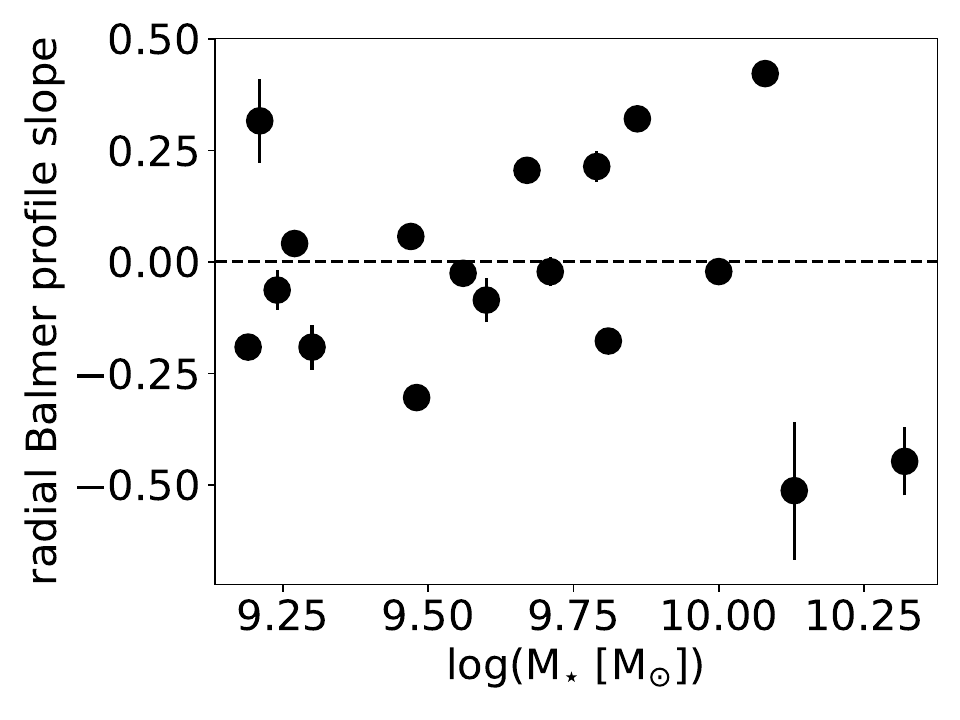}
    \includegraphics[width=0.45\linewidth]{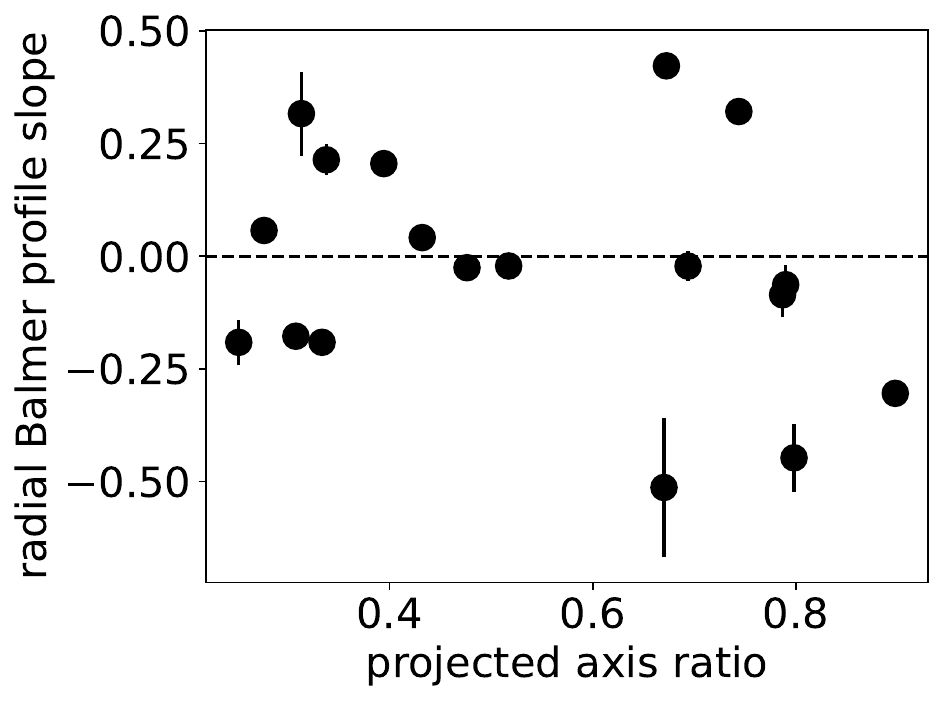}
    \caption{{\textbf{Left:}} Slope of the radial Balmer decrement profile (i.e., \Ha/\Hb\ vs. radius, in units of kpc$^{-1}$) 
    as a function of stellar mass. Positive (negative) values indicate larger (smaller) attenuation at larger radii. 
    Within the stellar mass range probed by our sample, galaxies exhibit a wide diversity of radial attenuation profiles, with no clear trend between profile slope and stellar mass. 
    {\textbf{Right:}} The radial Balmer decrement profile slope as a function of galaxy inclination, parametrized by the projected axis ratio (b/a). We do not observe a correlation between the radial attenuation gradient and the projected axis ratio, 
    indicating that the observed variation in radial Balmer decrement profiles is not driven by projection effects.
    }
    \label{fig:Bslope-mass}
\end{figure*}

\begin{figure*}[ht]
    \centering
    \includegraphics[width=0.45\linewidth]{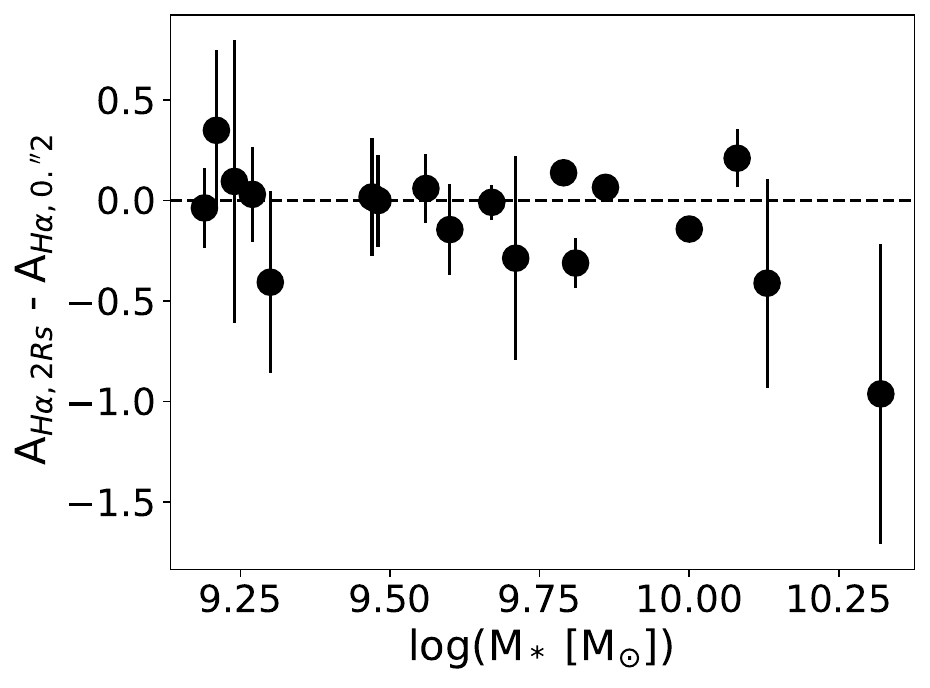}
    \includegraphics[width=0.45\linewidth]{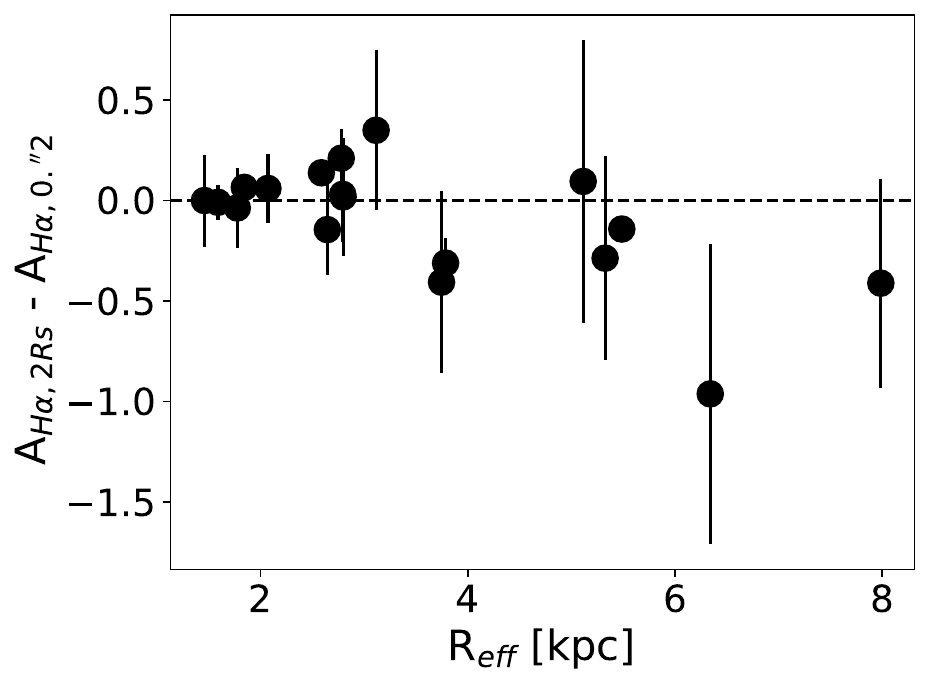}
    \caption{{\bf{Left:}} The difference between galaxy-wide and central attenuation as a function of stellar mass shows that galaxies at any given mass exhibit diverse radial attenuation profiles, from centrally concentrated (corresponding to negative values) to centrally depleted (positive values). {\bf{Right:}} The difference between galaxy-wide and central attenuation as a function of effective radius. Larger galaxies show stronger central attenuation relative to their outskirts.\\
    }
    \label{fig:dA-radius}
\end{figure*}

\begin{figure*}[ht]
    \centering
    \includegraphics[width=\linewidth]{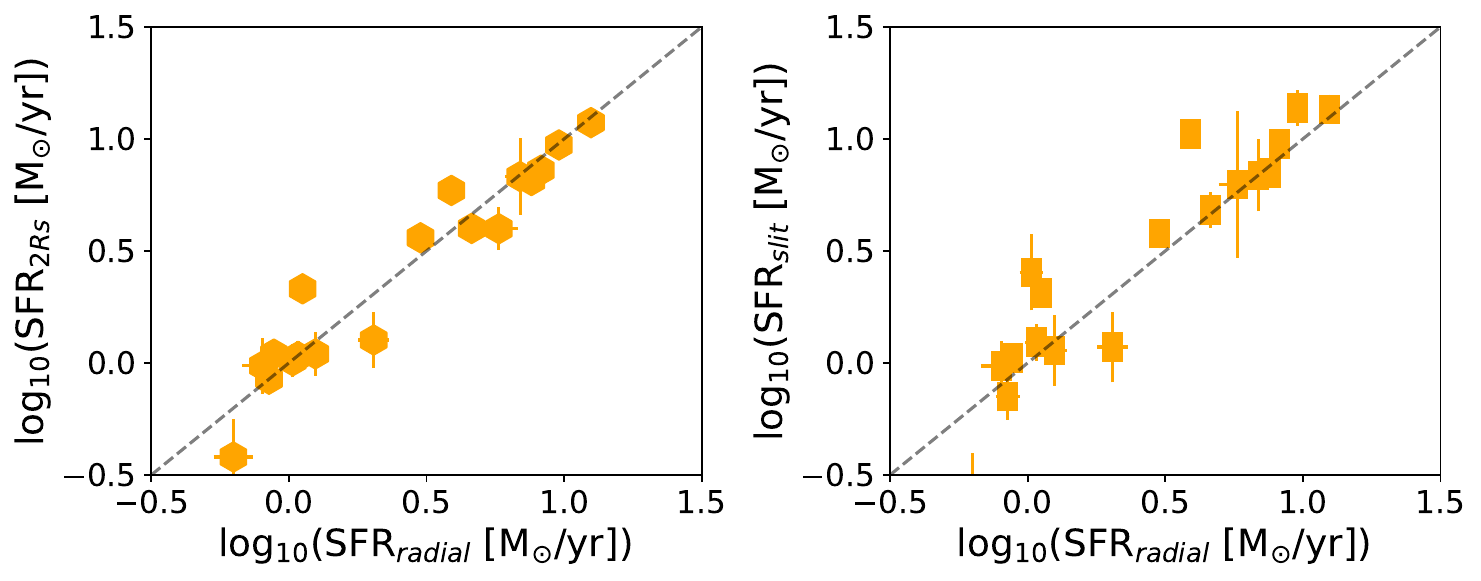}
    \caption{Comparison of star formation rates derived from \Ha\ flux within 2R$_S$, corrected using attenuation measured from a galaxy-wide aperture (2R$_S$; left panel) or a central aperture (0\farcs2; right panel), plotted against the cumulative SFR out to 2R$_S$ corrected using attenuation measured in 0\farcs2 annular radial bins. In both panels, the SFR estimates agree with each other, falling on the dashed 1-to-1 line to within 0.05~dex on average, 
    with RMSE values of $\sim$0.1~dex (left) and $\sim$0.2~dex (right).
    The lower scatter for the galaxy-wide correction indicates more robust integrated SFR estimates, while the central correction shows increased scatter due to variations in radial attenuation, but does not introduce a significant systematic offset.
    }
    \label{fig:sfr-comp}
\end{figure*}

\begin{table}[htbp]
\centering
\scriptsize 
\begin{tabular}{rrrrr}
\hline\hline
\textbf{ID} & $A(H\alpha)(2R_S)$ & $A(H\alpha)(0\farcs2)$ & $A(H\alpha)$(slit) & A(V) \\
\hline
2145 & 0.66 $\pm$ 0.14 & 0.70 $\pm$ 0.14 & 0.62 $\pm$ 0.15 & 0.53$_{-0.32}^{+0.29}$ \\
2465 & 0.31 $\pm$ 0.25 & 0.72 $\pm$ 0.38 & 0.35 $\pm$ 0.41 & 0.26$_{-0.19}^{+0.41}$ \\
3399 & 0.60 $\pm$ 0.09 & 0.91 $\pm$ 0.09 & 0.89 $\pm$ 0.10 & 0.46$_{-0.26}^{+0.21}$ \\
4391 & 0.40 $\pm$ 0.16 & 0.40 $\pm$ 0.16 & 0.20 $\pm$ 0.26 & 0.82$_{-0.21}^{+0.16}$ \\
6199 & 0.15 $\pm$ $\infty$ & 0.29 $\pm$ $\infty$ & 0.30 $\pm$ $\infty$ & 0.56$_{-0.17}^{+0.05}$ \\
6430 & 0.72 $\pm$ $\infty$ & 0.58 $\pm$ $\infty$ & 0.80 $\pm$ $\infty$ & 1.04$_{-0.09}^{+0.19}$ \\
7561 & 0.69 $\pm$ 0.32 & 0.34 $\pm$ 0.24 & 0.69 $\pm$ 0.29 & 0.66$_{-0.44}^{+0.34}$ \\
8365 & 0.18 $\pm$ 0.12 & 0.12 $\pm$ 0.12 & 0.13 $\pm$ 0.14 & 0.55$_{-0.26}^{+0.21}$ \\
8512 & 0.47 $\pm$ 0.25 & 1.44 $\pm$ 0.71 & 0.97 $\pm$ 0.85 & 0.77$_{-0.42}^{+0.32}$ \\
8576 & 0.46 $\pm$ 0.13 & 0.61 $\pm$ 0.18 & 0.67 $\pm$ 0.20 & 0.71$_{-0.11}^{+0.22}$ \\
8942 & 0.86 $\pm$ $\infty$ & 0.79 $\pm$ $\infty$ & 1.49 $\pm$ $\infty$ & 1.18$_{-0.13}^{+0.11}$ \\
10502 & 0.89 $\pm$ 0.45 & 1.30 $\pm$ 0.27 & 0.90 $\pm$ 0.42 & 0.77$_{-0.43}^{+0.20}$ \\
10863 & -0.56 $\pm$ 0.44 & -0.66 $\pm$ 0.54 & -1.19 $\pm$ 0.71 & 0.66$_{-0.40}^{+0.25}$ \\
11225 & 0.40 $\pm$ 0.06 & 0.41 $\pm$ 0.06 & 0.45 $\pm$ 0.06 & 0.61$_{-0.10}^{+0.19}$ \\
11944 & 0.48 $\pm$ 0.18 & 0.45 $\pm$ 0.15 & 0.64 $\pm$ 0.21 & 0.75$_{-0.39}^{+0.24}$ \\
12015 & 1.14 $\pm$ 0.12 & 0.93 $\pm$ 0.09 & 1.56 $\pm$ 0.20 & 0.88$_{-0.55}^{+0.29}$ \\
12071 & 0.59 $\pm$ 0.19 & 0.57 $\pm$ 0.22 & 1.58 $\pm$ 0.43 & 0.90$_{-0.18}^{+0.19}$ \\
29470 & -0.02 $\pm$ 0.33 & 0.27 $\pm$ 0.39 & -0.10 $\pm$ 0.41 & 0.59$_{-0.22}^{+0.23}$ \\
\hline
\end{tabular}
\caption{\Ha\ attenuation A(\Ha) measurements (in magnitudes) derived using three different apertures: galaxy-wide (2R$_S$), central (0\farcs2), and single slit. We also include A(V) values from SED fitting from the UVCANDELS catalog \citep{mehta24, wang25}.}
\label{tab:att_table}
\end{table}

\section{Discussion}
\label{sec:discussion}

\subsection{Radial attenuation profiles are diverse}
\label{sec:rad_profile}

We quantify the slope of the radial Balmer decrement profile of each galaxy by fitting a linear function to the Balmer decrement measurements in 0\farcs2 radial bins as a function of radius. A negative slope corresponds to a decreasing Balmer decrement with radius, i.e. higher attenuation in the center.
Within the MSA-3D IFS data-set, we find that galaxies at $z\sim1$ display a wide range of radial attenuation profiles (e.g., Figure~\ref{fig:radial_trends}). While we recover the known trend of increasing attenuation with higher stellar mass (Figure~\ref{fig:attenuation-mass}, top panels), Figure~\ref{fig:Bslope-mass} shows no clear trend in the slope of the radial Balmer decrement profile 
with stellar mass. Within our sample of 18 galaxies, six show significant negative A(\Ha) radial gradients and they span the full observed stellar mass range 
$\log{\mathrm{M_\star}/\Msun} = 9.1$--10.3. 
We can also examine radial gradients by comparing the integrated (within 2~R$_s$) and central (within 0\farcs2) attenuation as reported in Table~\ref{tab:att_table}. Their difference is shown in Figure~\ref{fig:dA-radius}. Galaxies with negative attenuation gradients have larger central A(\Ha) and thus negative $\mathrm{A_{2 R_s} - A_{0\farcs2}}$. The few most massive galaxies in our sample with $\log({\mathrm{M_\star}/\Msun}) > 10$ have negative gradients on average, but overall we find moderate scatter around the sample average value. While there is no clear trend with mass, we find a stronger trend with size such that galaxies with larger effective radii tend to have higher central attenuation (Pearson r~$=-0.63$, p~$=0.005$, slope~$=-0.105\pm0.033$, Figure~\ref{fig:dA-radius}, left panel). 
At low redshift, \citet{battisti26} report negative Balmer decrement gradients that steepen with stellar mass, on average, in the SAMI and MAGPI surveys ($z\lesssim0.4$). We do not find an analogous trend with stellar mass over a comparable mass range, which may be due to the diversity in attenuation gradients at a fixed mass, or simply due to the smaller size of our sample.

Beyond mass and size, we also examine whether the diversity in radial attenuation gradients could arise from projection effects. Although edge-on galaxies are expected to exhibit overall higher attenuation compared to face-on galaxies, due to longer dust path lengths \citep[e.g.][]{tuffs04, wild11, chevallard13, battisti17}, 
we find no significant correlation between the integrated attenuation at $2R_S$, $A(\mathrm{H}\alpha)(2R_S)$, and projected axis ratio ($b/a$; Pearson $r=-0.07$, $p=0.78$). Thus, edge-on and face-on galaxies in the MSA-3D sample exhibit comparable overall attenuation.
We likewise find no significant correlation between the radial Balmer decrement slope and inclination (projected axis ratio b/a; right panel of Figure~\ref{fig:Bslope-mass}). Galaxies with negative, flat and positive gradients span the full range of observed axis ratios. 
This lack of dependence of both the overall attenuation and radial gradient on inclination suggests that the attenuation is not strongly dominated by a diffuse dust component in the ISM and may instead be associated primarily with localized, clumpy star-forming regions, consistent with the inclination-independent attenuation found at cosmic noon by \citet{lorenz23}. 
The radial attenuation gradient therefore appears to reflect the intrinsic radial dust distribution relative to the stellar and ionized-gas components, rather than projection. 
The observed diversity in individual attenuation profiles may therefore reflect variations in evolutionary states of galaxies that are not captured by population-averaged measurements.

The scatter and weak trend with mass in our sample is complementary, and somewhat in contrast, to previous results based on stacking analyses. 
\cite{nelson2016_A} stacked emission line maps from HST grism spectra and found that more massive galaxies exhibit negative attenuation gradients and stronger central attenuation.  
While stacking increases signal-to-noise ratios and is effective for identifying average trends across galaxy populations, it does not provide information on the intrinsic diversity of individual galaxies. 
The diversity of attenuation profiles seen in MSA-3D may reflect differences in galaxies’ evolutionary states. Galaxies with negative gradients may be more evolved disks, e.g. having undergone fewer recent merger events, therefore having more time to settle and resulting in centrally concentrated dust distributions.
Conversely, those with flat or positive gradients may have experienced recent merger activity or turbulence, resulting in more uniform or even centrally depleted attenuation profiles.

The presence and diversity of attenuation gradients highlights the limitations of applying single-aperture based attenuation corrections for spatially extended galaxies. Relying on single-aperture measurements can result in over- or underestimation of the total dust content, which affects measurement of attenuation-corrected star-formation activity. 
Figure~\ref{fig:dA-radius} shows that the difference between a central aperture and galaxy-averaged A(\Ha) can be as large as $\sim$0.5 mag in some cases, with an RMS scatter of 0.3~mag.
Additionally, measurement of radial dust attenuation profiles is essential to accurately trace the spatial distribution of star formation activity. This is particularly relevant in systems undergoing significant morphological transformation, such as disk settling and in-situ bulge growth at $z \sim 1$.

\subsection{Total star formation rates}
\label{sec:totalSFR}

While the dust attenuation measured for an individual galaxy depends on the choice of aperture \citep[Table~\ref{tab:att_table}, Figure~\ref{fig:dA-radius};][]{barisic2025}, all apertures considered in this work yield generally consistent sample-average results for attenuation and star formation rate (Figure~\ref{fig:attenuation-mass}). We now examine the extent to which aperture effects introduce additional scatter in derived measurements. Essentially, larger scatter in radial attenuation profiles (Figure~\ref{fig:radial_trends}) causes a larger degree of aperture effects. 
Our most accurate estimate of properties such as the total attenuation-corrected SFR is from the radial bins, which account for overall radial structure of dust attenuation\footnote{Correcting individual resolution elements would be preferred, but is not practical given the limited signal/noise of the data.}. Figure~\ref{fig:sfr-comp} compares the total SFR from summed radial bins with that derived using a single attenuation based on the 2R$_S$ and single-slit apertures. In both cases the average difference in SFR is close to zero (mean difference within 0.05 dex), indicating little overall bias, as found previously (e.g., Figure~\ref{fig:attenuation-mass}). The difference between the 2R$_S$ and summed radial apertures furthermore has a small RMSE scatter of 0.12 dex. The intrinsic scatter is only 0.06 dex considering measurement uncertainties. In contrast, the single-slit aperture SFRs exhibit a larger 0.19 dex RMSE scatter which is predominantly intrinsic. This larger scatter for the single-slit case is explained by the diversity of radial attenuation gradient slopes, whereas the larger 2R$_S$ aperture more accurately captures the radially-averaged attenuation.

The results of Figure~\ref{fig:sfr-comp} highlight the spatial structure and resulting potential for aperture effects to create increased scatter in derived galaxy properties, despite recovering the same average trends such as the star forming main sequence (e.g., Figure~\ref{fig:attenuation-mass}). We find significant variation even at fixed stellar mass (e.g., Figure~\ref{fig:Bslope-mass}). 
This underscores the need for spatially resolved measurements to uncover the diversity of galaxy structure and evolutionary pathways, 
since different growth and quenching mechanisms leave distinct signatures in attenuation and star formation profiles. For instance, gas-rich mergers and gas compaction are expected to produce centrally concentrated attenuation and star formation \citep[e.g.,][]{dekel14,zolotov15,tacchella16}, while inside-out quenching can flatten/invert the attenuation profile as central star formation and dust are depleted \citep[e.g.][]{tacchella18,spilker19}. 
Strangulation predicts a more uniform decline in star formation and a flat/declining attenuation profile \citep[e.g.,][]{peng15}. 
The prevalence of flat/positive gradients in our sample does not suggest that the dust is centrally concentrated. However, we note that Balmer emission can miss the most dust-obscured star formation, while combining the Balmer decrement with a longer wavelength tracer (e.g. Pa$\alpha$) is a promising future avenue to reveal any such obscuration including in the central regions.

\subsection{Nebular vs. stellar reddening}
\label{sec:ebv_disc}

The comparison between stellar and nebular reddening provides further insight into the spatial distribution of dust in galaxies. 
Reddening E(B$-$V)$_{\rm star}$ derived from the stellar continuum traces dust in the line of sight of the stellar component, 
whereas E(B$-$V)$_{\rm gas}$ traces dust associated with ionized gas in \HII\ regions and is linked to ongoing star formation. Differences between these two measures provide a probe of how dust is distributed relative to the stellar and star-forming components.

\begin{figure}[t!]
    \centering
    \includegraphics[width=\linewidth]{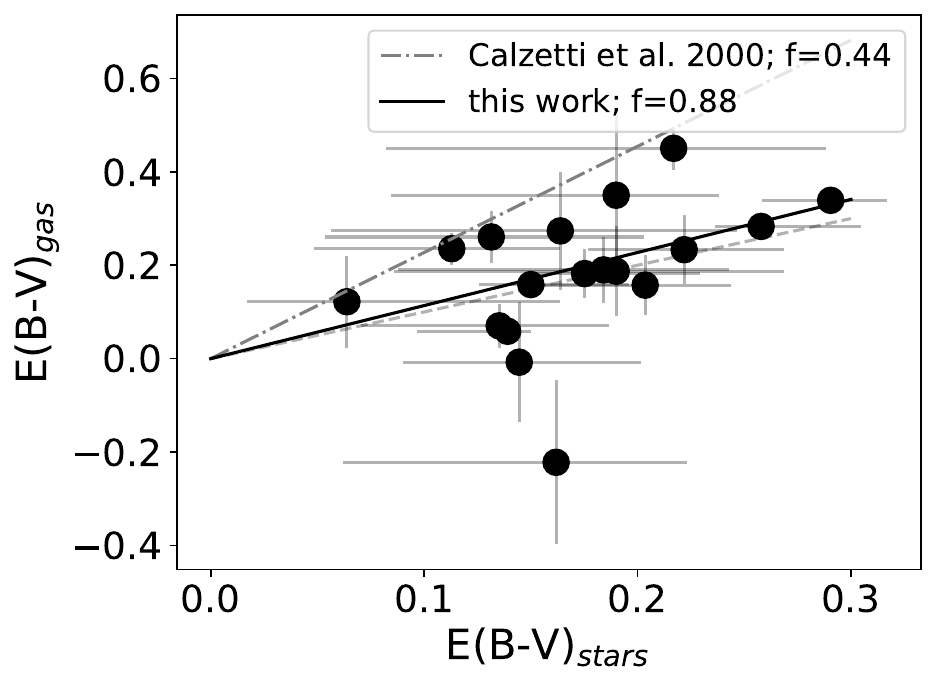}
    \caption{Comparison of stellar and nebular reddening (derived from the 2R$_{S}$ aperture), showing that stellar reddening is slightly lower than nebular reddening, with a median ratio of E(B-V)$_{\rm star}$ / E(B-V)$_{\rm gas}$ $\approx$ 0.88.
    }
    \label{fig:f-value}
\end{figure}

In the local universe, \cite{calzetti2000} found empirically  that nebular reddening exceeds stellar reddening, with a ratio of f = E(B$-$V)$_{\rm star}$ / E(B$-$V)$_{\rm gas}$ = 0.44 for a sample of 8 starburst galaxies (shown in Figure~\ref{fig:f-value}).  
Studies at $z\sim$1-2 report a variety of f-values. While studies such as \cite{buat18, mancini11, shivaei20, forster2009} find that nebular reddening is $\sim$2$\times$ the stellar reddening, consistent with the results based on local star-forming and starburst galaxies \citep[f$\sim$0.5, e.g.][]{calzetti2000, wild11, ly12, battisti16}, other studies find f-values closer to unity \citep[f$\sim$0.8-0.9, e.g.][]{pannella15, kashino13, puglisi16, reddy10, shivaei15a}. 
Part of the diversity in reported f-values can be attributed to differences in the assumed stellar attenuation curve \citep[e.g. Calzetti vs. SMC;][]{reddy20}, although several studies have also suggested that f-values vary with galaxy properties, including stellar mass, SFR, SFR surface density, sSFR, and galaxy size \citep[e.g.][]{price2014, reddy15, puglisi16, reddy20, shivaei20,song2026}. 
Ratios closer to unity are thought to indicate a more uniform dust geometry, where both stellar and nebular emission undergo similar reddening as a result of compact sizes and high SFR surface densities.

In our sample we find a median reddening ratio of f = $0.88^{+0.08}_{-0.37}$ using the integrated 2R$_S$ measurements, where uncertainties represent the interquartile 
range (see Figure~\ref{fig:f-value}). As described in Section~\ref{sec:attenuation}, we derive the nebular E(B-V)$_{\rm gas}$ based on the Balmer decrement, and the stellar E(B-V)$_{\rm star}$ using SED-based A(V) values from UVCANDELS (Table~\ref{tab:att_table}; \citealt{wang25}). 
We note that part of the scatter in reported f-ratios across the literature stems from differences in the assumed attenuation curves, which directly affects the inferred reddening: e.g. steeper curves yield smaller E(B$-$V)$_{\rm star}$ (lower f-values), 
whereas shallower curves produce the opposite trend \citep{reddy20}. The physical interpretation of f-value is further complicated by the fact that the nebular and stellar reddening curves are not necessarily the same. Because E(B$-$V)$_{\rm star}$ and E(B$-$V)$_{\rm gas}$ are derived using different attenuation curves, a similar E(B$-$V) for the two components does not imply a similar total attenuation \citep[see also][]{reddy20}. Comparisons of f-values between studies that adopt different stellar curves should therefore be interpreted with this caveat in mind. Reassuringly, the qualitative result that the nebular reddening generally equals or exceeds the stellar reddening is robust to the choice of stellar attenuation curve \citep{reddy20}.

\begin{figure*}[ht]
    \centering
    \includegraphics[width=0.45\linewidth]{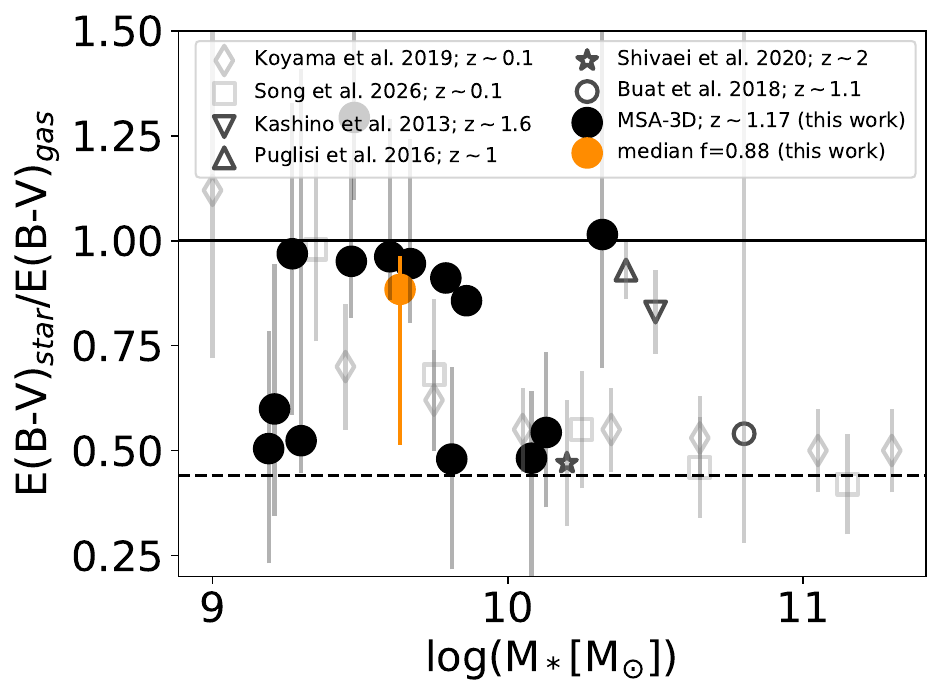}
    \includegraphics[width=0.45\linewidth]{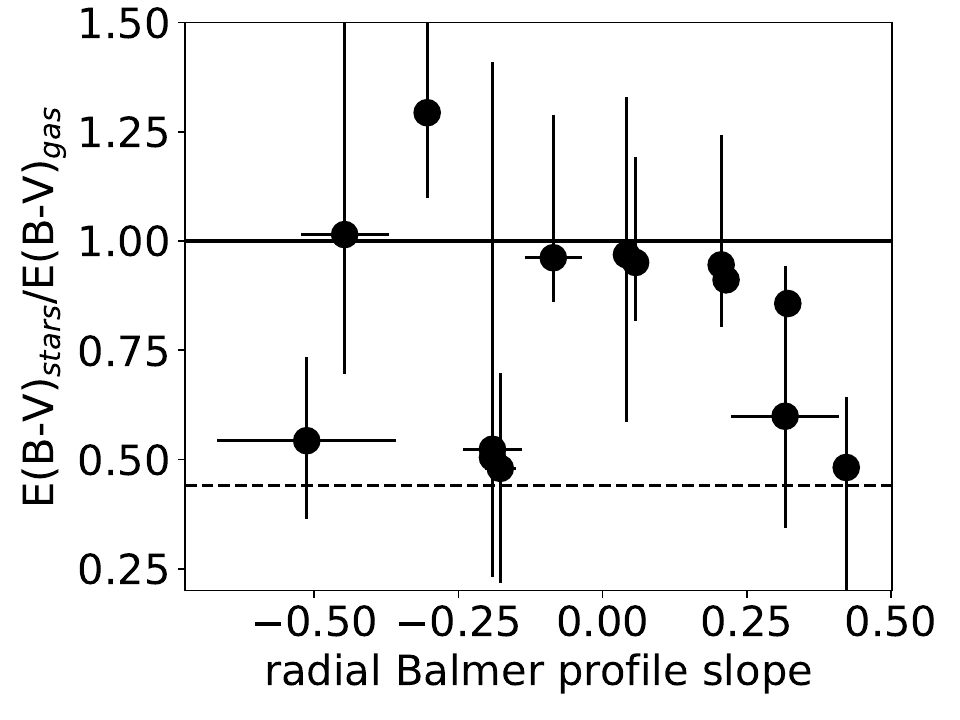}
    \caption{The stellar to nebular E(B-V) color excess ratio (f) as a function of stellar mass (left) and the slope of the radial Balmer decrement profile (right). Horizontal lines mark f$=$1 (solid) and \cite{calzetti2000} f$=$0.44 values. {{\bf{Left:}} Black circles show our MSA-3D sample, with the orange point indicating the median (and interquartile range). Comparison measurements are shown from \cite{kashino13}, \cite{puglisi16}, \cite{koyama19}, and \cite{song2026}. While local results show f decreasing with stellar mass, our sample shows no significant mass dependence over the range probed, with a median f=0.88 and substantial galaxy-to-galaxy variation. } 
    {{\bf{Right:}} We do not find significant correlation with the radial Balmer decrement profile slope, indicating that the integrated color excess ratio is not driven by the radial distribution of nebular attenuation within galaxies.}
    }
    \label{fig:ebv-mass}
\end{figure*}

We examine the dependence of f on stellar mass within our sample (Figure~\ref{fig:ebv-mass}). Some local studies find that f decreases with stellar mass, with f$\sim$1 at low masses evolving towards the \cite{calzetti2000} f$=$0.44 value at high masses \citep{koyama19, song2026}. As noted above, intermediate redshift studies at intermediate-to-high masses \citep[e.g.][]{kashino13, puglisi16, buat18, shivaei20} likewise report f-ratios ranging from near unity to that of \cite{calzetti2000}. Across the mass range probed by our sample we find no significant trend with stellar mass, with galaxy-to-galaxy variations around a median f$=$0.88, lying near the local f-mass relation (Figure~\ref{fig:ebv-mass}).

As discussed in Section~\ref{sec:rad_profile}, the radial attenuation profiles of galaxies in our sample are predominately flat or even positive, potentially indicating a spatially extended and clumpy dust distribution. 
As galaxies at $z\sim1$ are undergoing significant morphological evolution (from clumpy and turbulent, to settled thin disks), the combination of high SFRs, ongoing internal secular evolution, and recent mergers or feedback processes all may play a role in the emergent dust geometry, reflected in the observed f-values and the shape of radial attenuation profiles.

We therefore explore whether f depends on the spatial distribution of nebular attenuation, quantified by the slope of the radial Balmer decrement profile (Figure~\ref{fig:ebv-mass}). We find no significant correlation between f-values and the radial attenuation gradient. This is expected, as f is an aperture-integrated measure of stellar-to-nebular reddening, whereas the Balmer decrement slope characterizes the radial distribution of nebular attenuation. The lack of correlation therefore indicates that our integrated f measurements are not significantly affected by the radial attenuation distribution.
Overall we conclude that our sample generally exhibits higher attenuation of ionized gas relative to stars, with a spread in f values similar to the range seen in nearby galaxies.

\section{SUMMARY}
\label{sec:summary}

Using spatially resolved MSA-3D spectroscopy, we investigate the diversity of radial attenuation profiles in star-forming galaxies at z$\sim$1, and examine how measurements based on a single aperture affect 
the measured attenuation and resulting attenuation-corrected SFRs. 

We find that galaxies in our sample exhibit a diversity of radial attenuation profiles, with no strong dependence on stellar mass. While some galaxies show clear negative gradients consistent with centrally concentrated dust, many display flat or even positive gradients. This highlights the importance of spatially resolved measurements in capturing the full diversity of intrinsic dust distributions in individual galaxies. The variety of profiles may reflect differences in their respective evolutionary stage with respect to disk settling, merger history, or internal processes (e.g., feedback, secular evolution).

Our comparison of attenuation derived from central (0\farcs2) versus galaxy-wide (2R$_S$) flux measurements reveals that the measured attenuation can differ substantially depending on the choice of aperture. 
Despite these differences, SFRs corrected using attenuation from different apertures yield consistent sample-averaged values,  
with RMS scatters of 0.12~dex for the 2R$_S$ aperture and 0.19~dex for the single-slit aperture, relative to the summed radial aperture SFRs.
This indicates that despite significant variations in measured attenuation based on single apertures, there is no 
systematic bias in the recovered global SFRs across the sample, with mean offsets within $0.05$~dex relative to radially summed SFRs.

Finally, we examine the ratio of stellar-to-nebular reddening, finding a median value of f$=$0.88 (interquartile 
range f=0.51--0.96), indicating that the stellar and nebular components experience similar reddening. The near unity f-values, together with the prevalence of flat/positive radial attenuation profiles, suggest that dust distribution for many galaxies in our sample is clumpy and radially extended rather than strongly centrally concentrated. 

The diversity we find in radial attenuation profiles has implications for galaxy growth and structural evolution. For example, in situ bulge growth via dust-obscured star formation can be missed by a global attenuation estimate, while spatial mapping can reveal and correct for centrally concentrated dust attenuation. Similarly, signatures of inside-out growth and quenching can be identified from radial variations in attenuation-corrected SFR or sSFR. Furthermore we note that the Balmer decrement measurements used herein can underestimate the attenuation and SFR in highly obscured regions which are optically thick to \Ha. This can be addressed with future measurements of longer-wavelength Paschen lines (e.g., Pa$\alpha$, Pa$\beta$; \citealt{reddy2026a,reddy2026b}) in combination with the Balmer decrement. Already our Balmer-based results show a diversity of attenuation gradients which likely reflects the diverse evolutionary pathways — including both growth and quenching modes — of moderately massive galaxies at $z\sim1$. Mapping the star formation profiles with careful accounting for the spatially varying attenuation is thus a promising approach to understand the growth and morphological transformation of galaxies at $z\sim1$ as they settle onto the modern Hubble sequence.

\section*{ACKNOWLEDGEMENTS}

This work is based on observations made with the NASA/ESA/CSA James Webb Space Telescope. 
The data were obtained from the Mikulski Archive for Space Telescopes at the Space Telescope Science Institute, which is operated by the Association of Universities for Research in Astronomy, Inc., under NASA contract NAS 5-03127 for JWST. 
The specific observations analyzed can be accessed via \dataset[doi: 10.17909/s8wp-5w10]{https://doi.org/10.17909/s8wp-5w10}.
These observations are associated with program JWST-GO-2136. We acknowledge financial support from NASA through grant JWST-GO-2136. 
This work made use of observations and catalogs from the 3D-HST Treasury Program (GO 12177 and 12328) with the NASA/ESA Hubble Space Telescope, which is operated by the Association of Universities for Research in Astronomy, Inc., under NASA contract NAS5-26555.
XW and MJ are supported by the National Natural Science Foundation of China (grant 12373009), the CAS Project for Young Scientists in Basic Research Grant No. YSBR-062, the Fundamental Research Funds for the Central Universities, the Xiaomi Young Talents Program, and the science research grant from the China Manned Space Project. 
CAFG was supported by NSF through grants AST-2108230 and AST-2307327; by NASA through grant 21-ATP21-0036; and by STScI through grant JWST-AR-03252.001-A. J.M.E.S. acknowledges financial support from the European Research Council (ERC) Advanced Grant under the European Union’s Horizon Europe research and innovation programme (grant agreement AdG GALPHYS, No. 101055023).
Views and opinions expressed are, however, those of the author(s) only and do not necessarily reflect those of the EU or the ERC. Neither the EU nor the granting authority can be held responsible for them. T.T. is supported by the JSPS Grant-in-Aid for Research Activity Start-up
(25K23392) and the JSPS Core-to-Core Program (JPJSCCA20210003).
AA acknowledges support from the INAF Large Grant 2022 “Extragalactic Surveys with JWST” (PI Pentericci) and from the European Union – NextGenerationEU RFF M4C2 1.1 PRIN 2022 project 2022ZSL4BL INSIGHT.

\software{JWST \citep{jwst}, MSA3D \citep{msa3d}, Scipy \citep{scipy}, Numpy \citep{numpy}, Astropy \citep{astropy:2013, astropy:2018, astropy:2022}}

\facilities{JWST (NIRSpec MSA)}

\appendix

\section{Radial attenuation profiles of the sample}
\label{sec:attenuation_gradients_fullsample}

Radial attenuation profiles A(\Ha) and A(\Hb) for the full sample are shown in Figure~\ref{fig:attenuation_profiles_fullsample}.

\begin{figure}[htbp]
\centering

\begin{subfigure}{0.3\textwidth}
\includegraphics[width=\linewidth]{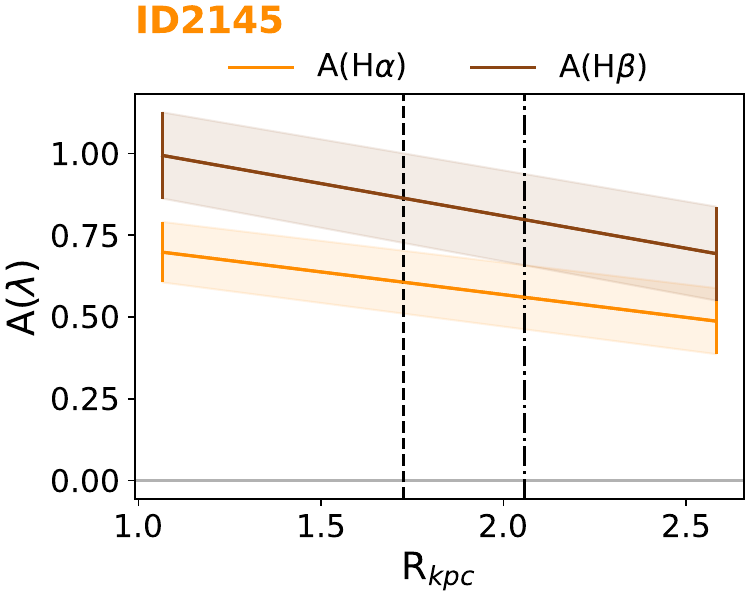}
\end{subfigure}
\begin{subfigure}{0.3\textwidth}
\includegraphics[width=\linewidth]{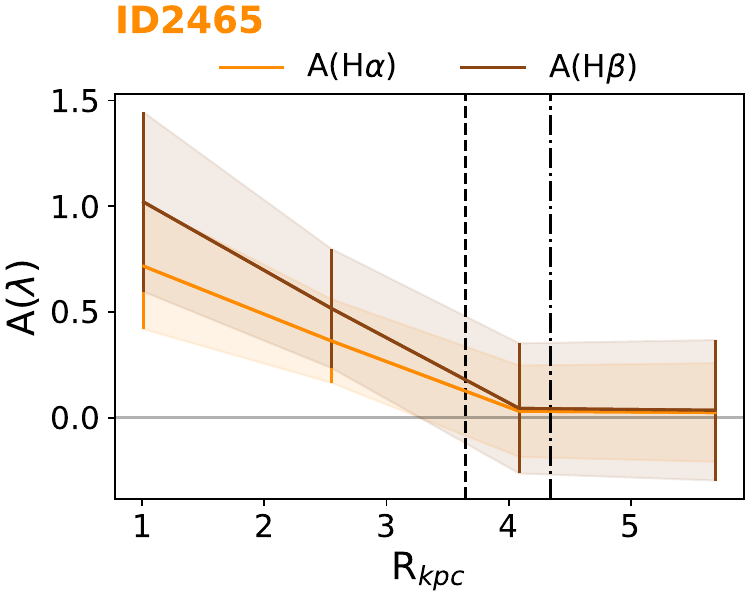}
\end{subfigure}
\begin{subfigure}{0.3\textwidth}
\includegraphics[width=\linewidth]{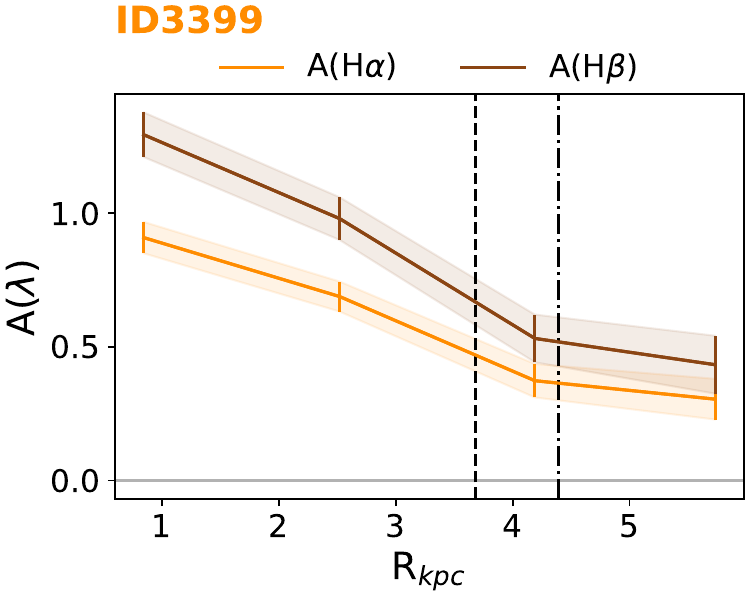}
\end{subfigure}

\medskip

\begin{subfigure}{0.3\textwidth}
\includegraphics[width=\linewidth]{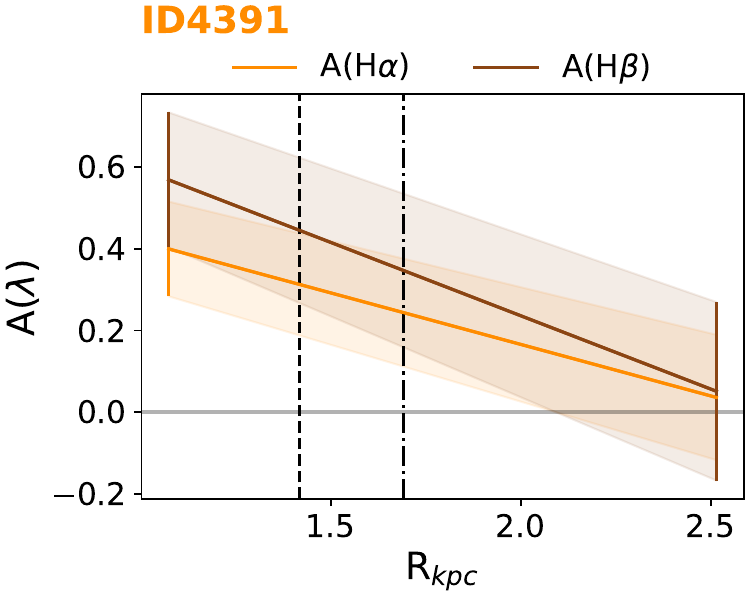}
\end{subfigure}
\begin{subfigure}{0.3\textwidth}
\includegraphics[width=\linewidth]{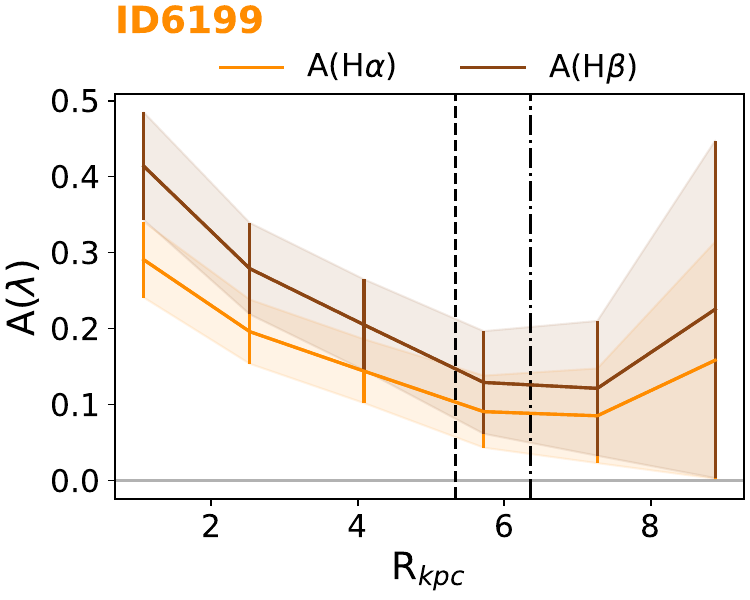}
\end{subfigure}
\begin{subfigure}{0.3\textwidth}
\includegraphics[width=\linewidth]{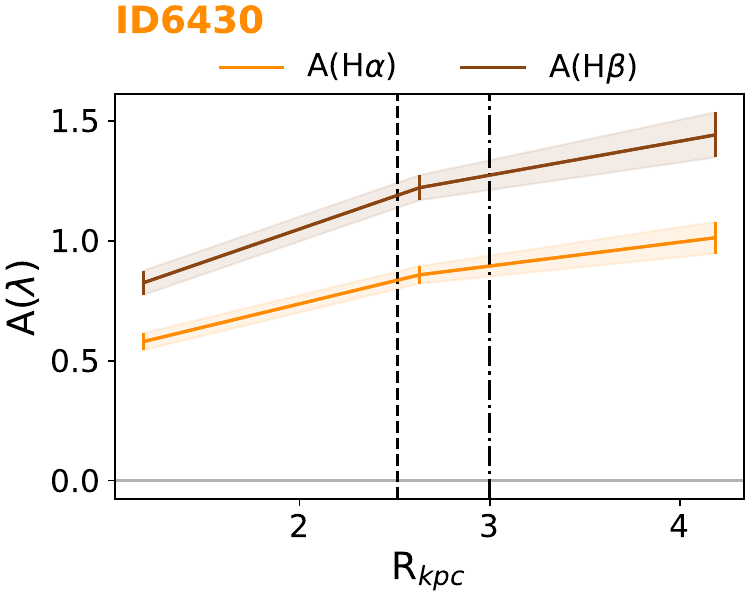}
\end{subfigure}

\medskip

\begin{subfigure}{0.3\textwidth}
\includegraphics[width=\linewidth]{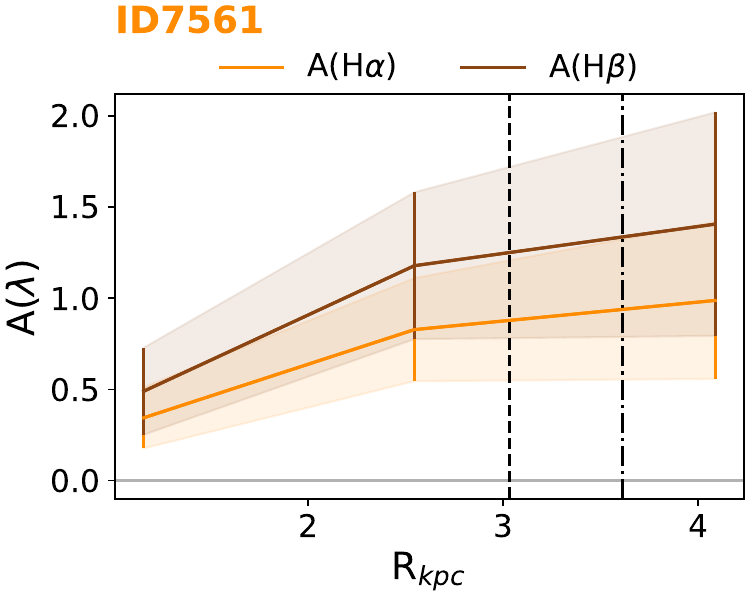}
\end{subfigure}
\begin{subfigure}{0.3\textwidth}
\includegraphics[width=\linewidth]{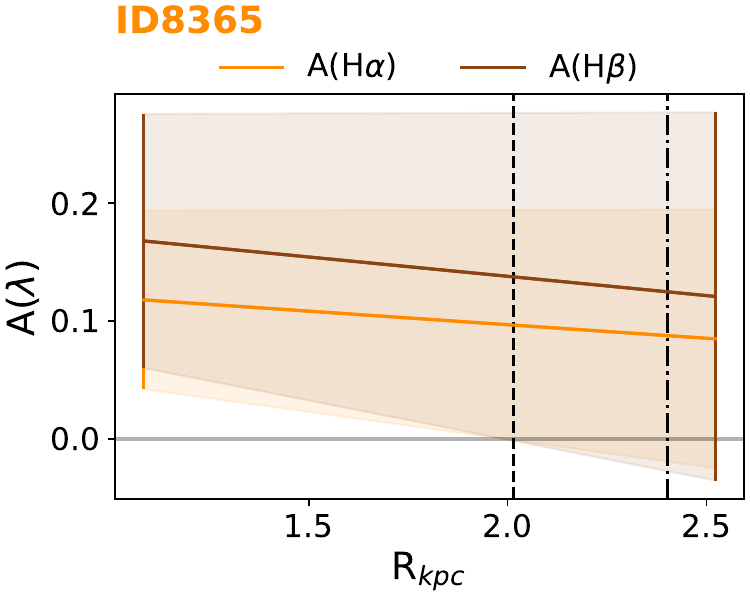}
\end{subfigure}
\begin{subfigure}{0.3\textwidth}
\includegraphics[width=\linewidth]{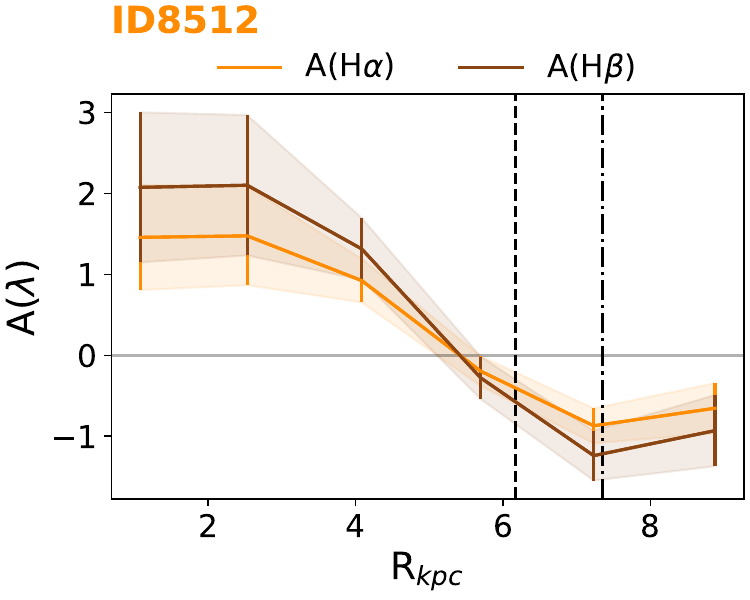}
\end{subfigure}

\medskip

\begin{subfigure}{0.3\textwidth}
\includegraphics[width=\linewidth]{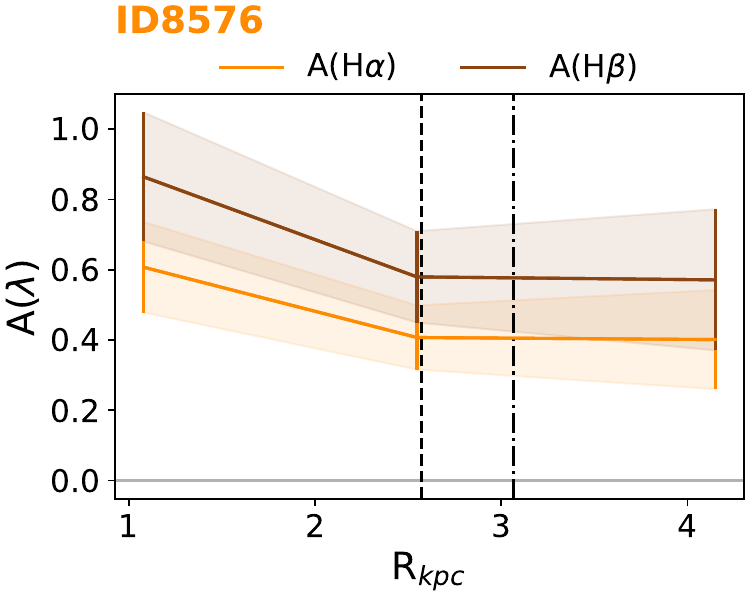}
\end{subfigure}
\begin{subfigure}{0.3\textwidth}
\includegraphics[width=\linewidth]{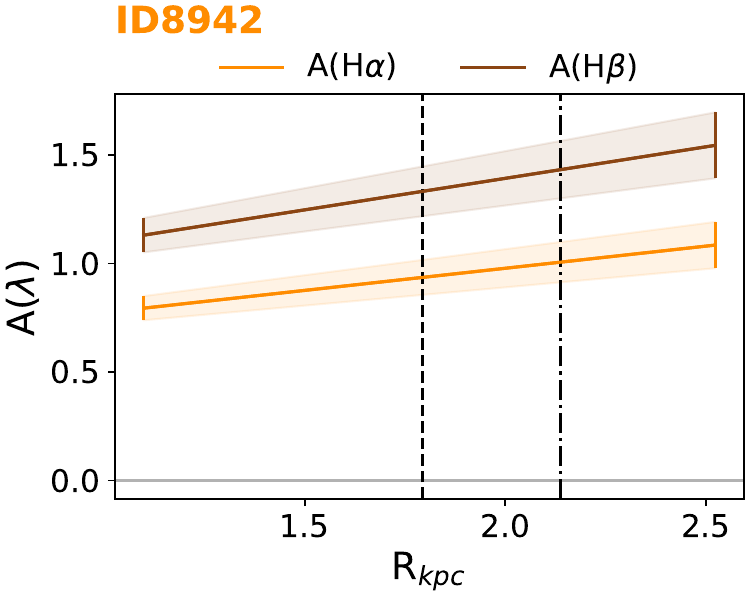}
\end{subfigure}
\begin{subfigure}{0.3\textwidth}
\includegraphics[width=\linewidth]{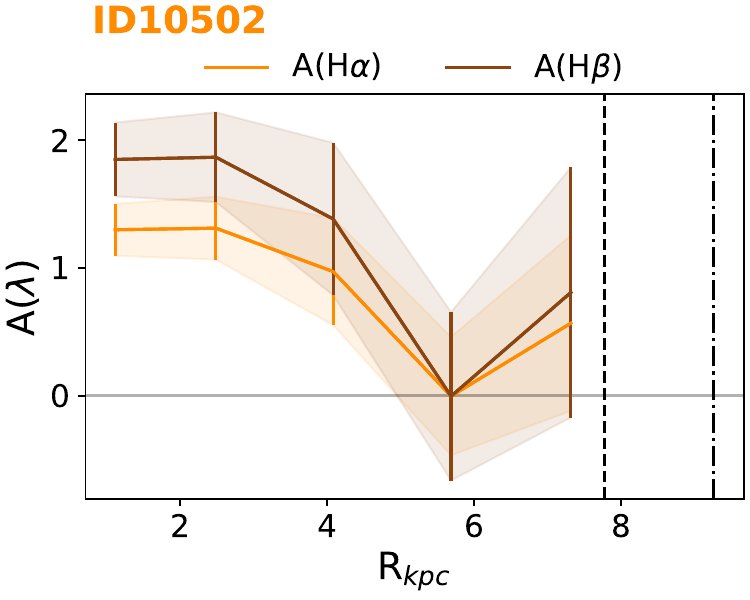}
\end{subfigure}

\caption{Radial attenuation profiles A(\Ha) (orange) and A(\Hb) (brown) for the full sample, derived from the Balmer decrement measured in each 0\farcs2 radial bin. Shaded regions indicate 1$\sigma$ uncertainties propagated from the Balmer decrement measurement errors in each bin. The vertical dashed and dash-dotted lines denote the effective radius (R$_{eff}$) and two scale radii (2R$_S$), respectively. The panels show the diversity of attenuation gradients across the sample.}
\label{fig:attenuation_profiles_fullsample}

\end{figure}

\begin{figure}[htbp]\ContinuedFloat
\centering

\begin{subfigure}{0.3\textwidth}
\includegraphics[width=\linewidth]{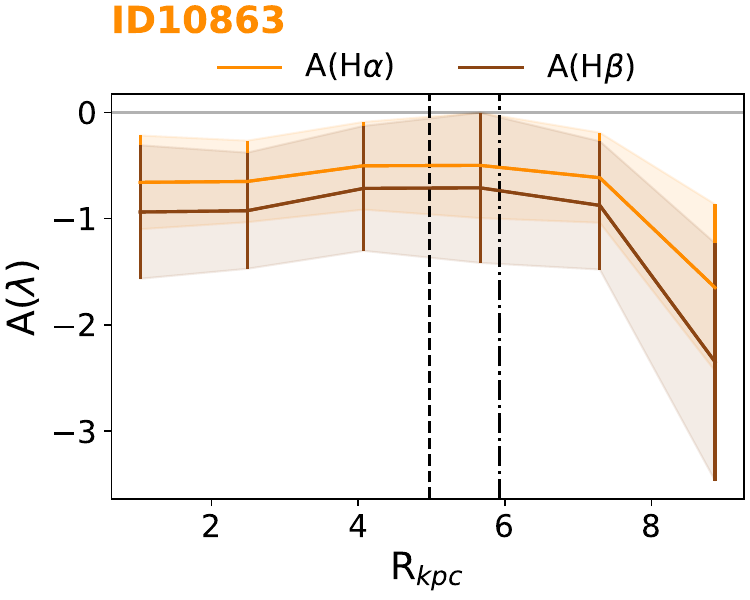}
\end{subfigure}
\begin{subfigure}{0.3\textwidth}
\includegraphics[width=\linewidth]{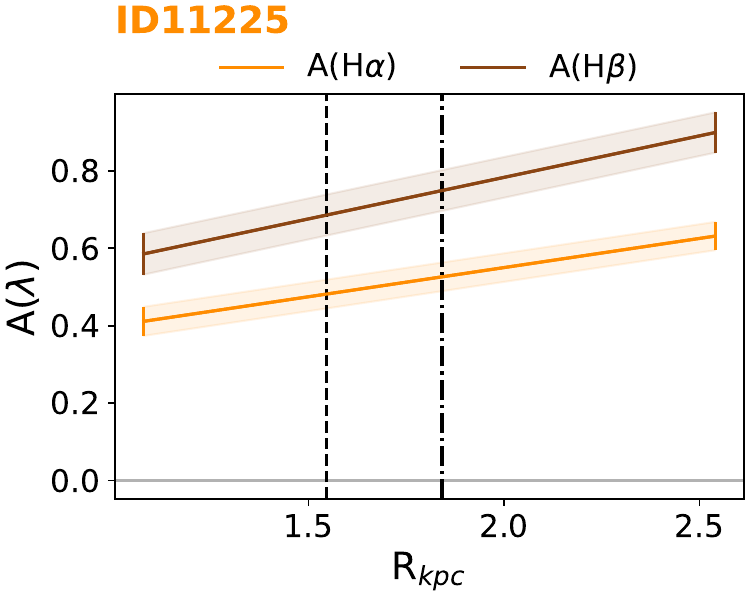}
\end{subfigure}
\begin{subfigure}{0.3\textwidth}
\includegraphics[width=\linewidth]{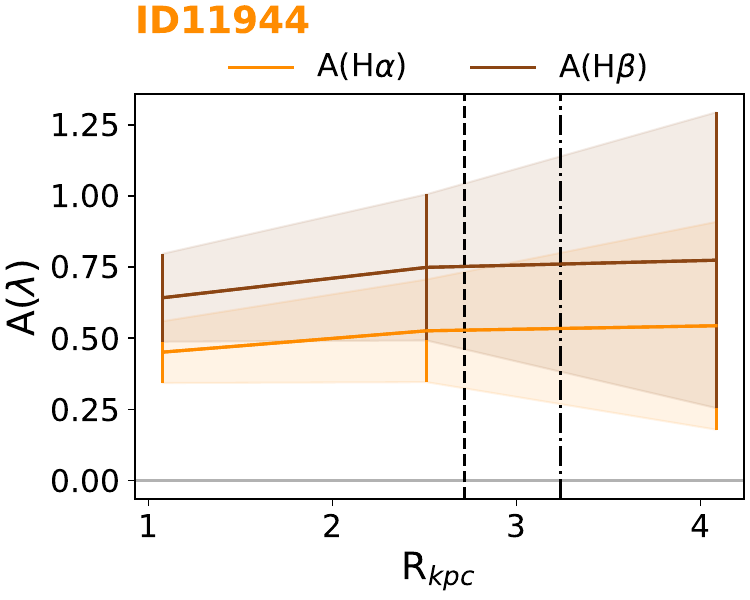}
\end{subfigure}

\medskip

\begin{subfigure}{0.3\textwidth}
\includegraphics[width=\linewidth]{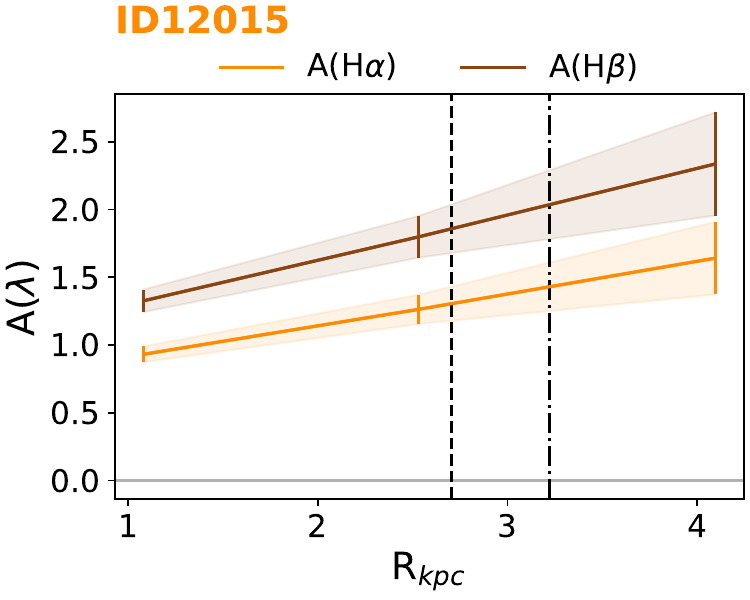}
\end{subfigure}
\begin{subfigure}{0.3\textwidth}
\includegraphics[width=\linewidth]{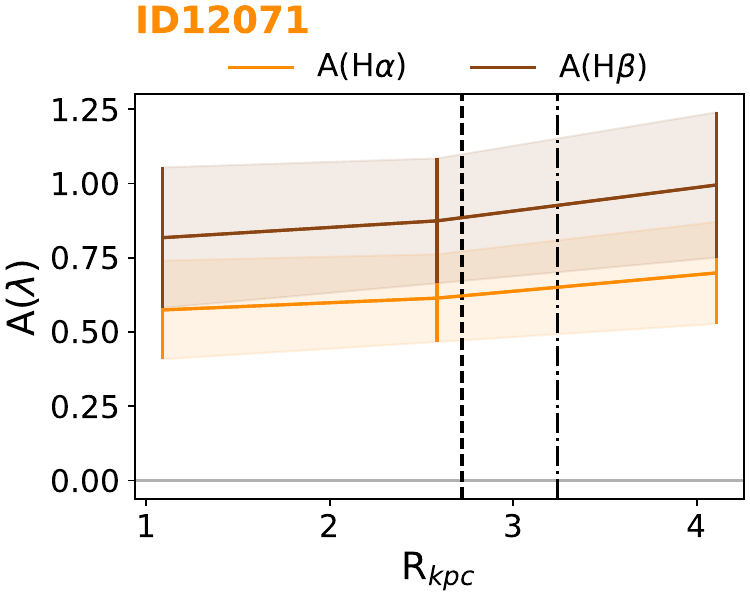}
\end{subfigure}
\begin{subfigure}{0.3\textwidth}
\includegraphics[width=\linewidth]{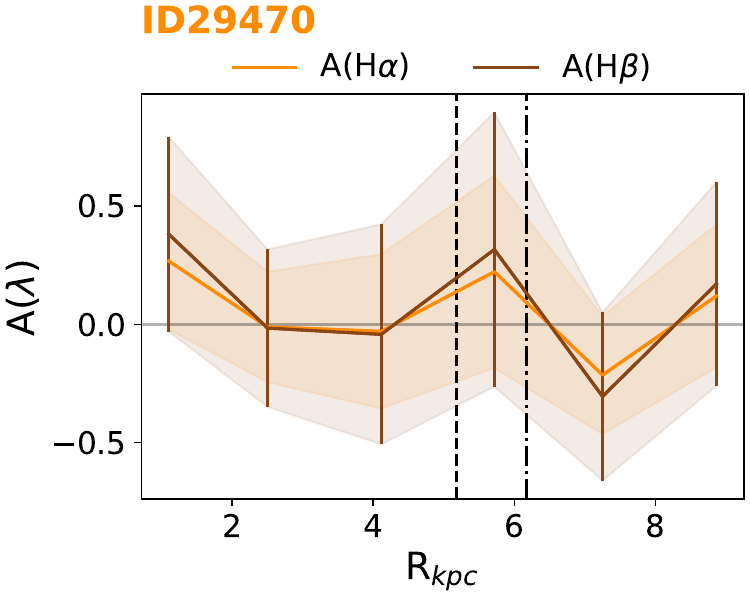}
\end{subfigure}

\caption{(continued)}

\end{figure}

\newpage
\bibliography{JWST_Paper}


\end{document}